\documentclass[usenatbib]{mnras}

\usepackage{newtxtext,newtxmath}
\usepackage{xspace,amsmath}
\usepackage{color,epsfig,amssymb,lscape}
\usepackage{array,colortbl}
\usepackage{url}

\usepackage[T1]{fontenc}
\usepackage{ae,aecompl}

\newcommand{\hii}{\relax \ifmmode {\mbox H\,{\scshape ii}}\else H\,{\scshape ii}\fi}
\newcommand{\mi}{\relax \ifmmode {\mu{\mbox m}}\else $\mu$m\fi}
\newcommand{\ha}{\relax \ifmmode {\mbox H}\alpha\else H$\alpha$\fi}
\newcommand{\hb}{\relax \ifmmode {\mbox H}\beta\else H$\beta$\fi}

\newcommand{\sii}{\relax \ifmmode {\mbox S\,{\scshape ii}}\else S\,{\scshape ii}\fi}
\newcommand{\siii}{\relax \ifmmode {\mbox S\,{\scshape iii}}\else S\,{\scshape iii}\fi}
\newcommand{\nii}{\relax \ifmmode {\mbox N\,{\scshape ii}}\else N\,{\scshape ii}\fi}
\newcommand{\oi}{\relax \ifmmode {\mbox O\,{\scshape i}}\else O\,{\scshape i}\fi}
\newcommand{\oii}{\relax \ifmmode {\mbox O\,{\scshape ii}}\else O\,{\scshape ii}\fi}
\newcommand{\oiii}{\relax \ifmmode {\mbox O\,{\scshape iii}}\else O\,{\scshape iii}\fi}
\newcommand{\neiii}{\relax \ifmmode {\mbox Ne\,{\scshape iii}}\else Ne\,{\scshape iii}\fi}

\newcommand{\rdostres}{\relax \ifmmode {\,\mbox{R}}_{\rm 23}\else \,\mbox{R}$_{\rm 23}$\fi} 

\newcommand{\ciii}{\relax \ifmmode {\mbox O\,{\scshape iii}}\else C\,{\scshape iii}\fi}
\newcommand{\civ}{\relax \ifmmode {\mbox O\,{\scshape iii}}\else C\,{\scshape iv}\fi}
\newcommand{\heii}{\relax \ifmmode {\mbox He\,{\scshape ii}}\else He\,{\scshape ii}\fi}

\newcommand{\gsim}{\hbox{\rlap{\lower.55ex\hbox{$\sim$}} \kern-.3em
\raise.4ex \hbox{$>$}}}
\newcommand{\lsim}{\hbox{\rlap{\lower.55ex\hbox{$\sim$}} \kern-.3em
\raise.4ex \hbox{$<$}}}

\usepackage{graphicx}	
\usepackage{amsmath}	
\usepackage{amssymb}	

\title[Model-based abundances in Type-2 AGNs]{A bayesian-like approach to
derive chemical abundances in Type-2 Active Galactic Nuclei based on photoionization models}

\author[P{\'e}rez-Montero et al.]{
E.\ P{\'e}rez-Montero$^{1}$\thanks{E-mail: epm@iaa.es},
O.\ L.\ Dors Jr.$^2$, J.\ M.\ V\'\i lchez$^1$, R.\ Garc\'\i a-Benito$^1$, M.\ V.\ Cardaci$^{3,4}$ \and G.\ F.\ H\"agele$^{3,4}$ \\
$^{1}$Instituto de Astrof\'\i sica de Andaluc\'\i a. CSIC. Apartado de correos 3004. 18080, Granada, Spain.\\
$^2$ Universidade do Vale do Para\'iba, Av. Shishima Hifumi, 2911, Cep
12244-000, S\~ao Jos\'e dos Campos, SP, Brazil\\ 
$^3$ Instituto de Astrof\'isica de La Plata (CONICET-UNLP), Argentina. \\
$^4$ Facultad de Ciencias Astron\'omicas y Geof\'{\i}sicas, Universidad Nacional de La Plata, Paseo del Bosque s/n, 1900 La Plata, Argentina.\\
}

\date{Accepted XXX. Received YYY; in original form ZZZ}

\pubyear{2017}

\begin{document}
\label{firstpage}
\pagerange{\pageref{firstpage}--\pageref{lastpage}}
\maketitle

\begin{abstract}
We present a new methodology for the analysis of the emission lines of the interstellar medium
in the Narrow Line Regions around type-2 Active Galactic Nuclei. 
Our aim is to provide a recipe that can be used for large samples of objects in a consistent way using different sets of optical emission-lines that takes into the account possible variations from the O/H - N/O relation to use [\nii] lines.
Our approach consists of a bayesian-like comparison between
certain observed emission-line ratios sensitive to total oxygen abundance, nitrogen-to-oxygen ratio and ionization parameter with the predictions
from a large grid of photoionization models calculated under the most usual conditions in this environment.
We applied our method to a sample of Seyfert 2 galaxies with optical emission-line fluxes and
determinations of their chemical properties from detailed models in the literature. Our results
agree within the errors  with other  results and confirm the high metallicity of the objects of the sample,
with N/O values consistent wit a large secondary production of N, but with a large dispersion. The obtained ionization parameters for
this sample are much larger than those for star-forming object at the same metallicity.

\end{abstract}




\section{Introduction}

Active galactic nuclei (AGNs) are among the most luminous objects in the Universe.
The intense and energetic radiation coming from the accretion disk and the jets around supermassive black holes in the centers of galaxies
is absorbed by the surrounding interstellar medium (ISM) and partially reemitted as very bright and prominent emission-lines that
contain valuable information about the physical conditions, the chemical abundances and the mechanical properties of
the gas under these very extreme conditions. Since these objects can be studied up to very high redshifts the correct
characterization of this kind of spectra can thus provide information about the cosmic evolution of galaxies.

It is widely accepted from the works by several authors (e.g. \citealt{ferland83, halpern83})
that, according to models, the main mechanism of ionization in the Narrow Line Region (NLR) in AGNs is photoionization.
This can in principle opens the gate to the estimation of the physical properties
and the ionic abundances in the gas by the measurement of the most prominent optical
emission-lines. However, it is known that the most widely used recipes to derive the total metallicity using this information
in star-forming regions (i.e. the $T_e$ method) leads to sub-solar metallicities in AGNs,
which are very low  as compared to the predictions from photoionization models 
or the expected values for their radial positions in galactic disks (e.g. \citealt{dors15}).

In this way models become a crucial tool to interpret the observed optical and ultraviolet  spectra of NLRs and they have been traditionally used
to provide calibrations for the derivation of total metallicity from the measurement of the emission lines (e.g. \citealt{storchi98,nagao06,dors14}).
Models are also used to derive abundances in star-forming regions, where it is known that
the $T_e$ method leads to precise abundance estimations. This occurs in the case that the
electron temperature cannot be derived because the required auroral lines (e.g. [\oiii] $\lambda$4363 \AA) is
too weak to be measured or owing to a too restricted observed spectral range.
Although many authors point out that models can lead to systematically higher oxygen abundances than those calculated
using the $T_e$ method also in the case of star-forming regions (e.g. \citealt{kd02, blanc15,asari16}),
these differences are much lower than in the case of the NLRs of AGNs and are even negligible in some works (e.g. \citealt{pm10,dors11}).

Models are also useful to overcome the problem of relative variations between different elemental 
abundances whose emission-lines are used as tracers of the total metallicity.
This is the case of the nitrogen-to-oxygen ratio (N/O), which has a monotonically growing behavior with
O/H in the high-metallicity regime (i.e. $Z \geq 0.3\cdot Z_{\odot}$) as predicted by galactic chemical evolution models under the assumption of total isolation,
as most of nitrogen production is secondary \citep{henry2000}.
However it is known that hydrodynamical effects can affect the ratio between secondary and primary elemental abundances
\citep{edmunds90}, what can cause non-negligible deviations in the derivation of O/H
or in the AGN identification  using N emission lines
such as [\nii] $\lambda$6583 \AA\ \citep{pmc09}.
This emission-line is especially bright and used in the case of the NLRs and it has been proposed as a direct estimator of
the total metallicity from the ratio of its flux with the  \ha\ one (e.g. \citealt{storchi98, groves06}) or in relation to the [\oii] $\lambda$3727 \AA\ \citep{castro17}, which
has a lower dependence on excitation.

In this work we propose a new method based on photoionization models
to derive chemical abundances in the NLRs of AGNs,
whose validity has been already proved for star-forming regions.
The code {\sc Hii-Chi-mistry} (hereafter {\sc HCm}, \citealt{hcm14}) establishes a
bayesian-like comparison between the relative observed optical line fluxes emitted by the ionized gas and the predictions from a grid of photoionization models covering a large range of input conditions in O/H, N/O and ionization
parameter ($U$).
The code, as probed for star-forming regions, has the advantages that i) firstly estimates the N/O ratio as a free parameter
so the [\nii] lines can be used as direct tracers of metallicity, ii) it is totally consistent with the predictions from the $T_e$ method, even in
the absence of any auroral line, and iii) provides consistent solutions regardless of the input emission-lines
given to find a solution, what makes this tool especially useful to compare sets of observations with different 
spectral ranges or redshifts.
The use of such a tool in the case of AGNs would allow to provide solutions to the estimation of
O/H and N/O for large samples of objects, in an automatic way and
supplying uncertainties, that can later be compared with other AGNs or star-forming galaxies
 in a consistent way..
Similar methodology than the one considered here  
 was recently used by 
\cite{thomas19} and \cite{mignoli19})
in order to calculate the metallicity of NLRs of 
local Seyfert 2 ($z \, \la \, 0.3$) and of 
obscured (type2) AGNs ($1.4 \, \la \, z \, \la \, 3.0$),
respectively.

The paper is organized as follows. In Section 2 we describe the observational sample
compiled and analyzed in \cite{dors17} and used here to test the accuracy of our method
as these objects have previous similar model-based estimations both for O/H and N/O using optical emission lines
 that can help to quantify the accuracy and uncertainty of our method.
In Section 3 we describe the {\sc HCm} code and the photoionization models  to be used as grid of values for it in the specific landscape of the NLRs of type-2 AGNs. 
In Section 4 we discuss the results obtained from the application of the code to the
control sample, the consistency of our method with the $T_e$ method and how our results vary as
a function of different input parameters in the models, such as the electron density or the shape of the ionizing spectral energy distribution. Finally, in Section 5 we summarize our results and present our conclusions.

\section{Description of the control sample}

In order to establish comparisons between the predicted abundances from our code 
with other model-based results in
a sample of objects with the required spectral information
we resorted to the objects described and analyzed in \cite{dors17}.
This sample consists of 47 Seyfert 1.9 and 2 galaxies 
at a redshift $z$ \la 0.1 with observations in the optical range covering all the required 
 reddening corrected, relative to \hb emission-line fluxes
to be used as an input by the code, including [\oii] $\lambda$3727 \AA, [\neiii] 3868 \AA, [\oiii] 4363 and 5007 \AA\AA,
[\nii] $\lambda$6583 \AA, and [\sii] $\lambda$$\lambda$6717,6731 \AA\AA, with a full width half maximum
lower than 1\,000 km $\cdot$ s$^{-1}$.

Despite these lines were measured using different observational techniques and apertures, all of them can be
unambiguously attributed to the NLR in each galaxy and thus can be considered
as an integrated emission that can be later reproduced in a single photoionization model.

We also used as control sample the abundance estimations
derived by \cite{dors17}. These authors used the {\sc Cloudy}
code to build detailed tailored photoionization models to reproduce observed optical 
narrow emission line intensities  compiled from the literature
in order to obtain quantitative determinations of oxygen and nitrogen 
abundances
for a sample of 44 AGNs. These authors found 
oxygen and nitrogen abundances in the range $0.4-2$ and $0.3-7.5$
times the solar values, respectively, concluding that
these galaxies have abundances quite similar to those of high-metallicity extragalactic \hii\ regions.

The list of reddening corrected emission lines, the control abundances and the names and redshifts of these galaxies can be found in \cite{dors17}.

\section{Description of the method}

The strategy followed to derive ionic abundances and ionization parameters described in this paper is rather similar
to that described in \cite{hcm14} to obtain these quantities in star-forming \hii\ regions. The idea is to establish a direct  comparison between a grid of
photoionization models covering a large range of input properties and a set of reddening-corrected emission lines relative
to a recombination {\sc Hi} line by means of a bayesian-like approach that later provides us with the most probable
values and their corresponding uncertainties. This method can be applied to a large number of objects using the same procedure without 
need of performing detailed modelling of the gas.

\subsection{The grid of models}

The grid of models used to be compared with a set of observed optical emission lines in type-2 AGNs was calculated by using the
code {\sc Cloudy} v.17.01 {\citep{cloudy}. This code calculates the emergent spectrum emitted by a spherical
gas distribution surrounding a central point-like ionizing source. We assumed that the gas was distributed homogeneously with
a filling factor of 0.1 and a constant density of 500 cm$^{-3}$, typical in the NLRs around type-2 AGNs \citep{dors14}. 
about the maximum value found by \cite{dors14}. The 
Spectral Energy Distribution (SED) was considered to be composed by two components: one representing the Big Blue Bump peaking at 1 Ryd, and the other a power law with spectral index $\alpha_x$ = -1 representing the non-thermal X-rays radiation. As usual, the continuum between 2 keV and 2500 \AA\ is described by a power law with a spectral index $\alpha_{ox}$=-0.8. 
This value represents the highest in the range derived in the sample of radio-intermediate and radio-loud quasars studied by \cite{miller11}, but it is the
most adequate to reproduce [\oiii]/\hb\ most type-2 AGNs according to \cite{dors17}.
The stopping criterion to measure the resulting emergent spectrum is that the proportion of free electrons in the ionized gas is lower than 98\%.
We considered a dust-to-gas ratio with the standard Milky Way value.
All chemical abundances were scaled to oxygen following the solar proportions given by \cite{asplund09}, with the
exception of nitrogen, that was left as an extra free input parameter in the models. 

We have chosen the input parameters in the grid of models according to the most usual conditions
observed in the NLRs of AGNs but additional uncertainties are expected in the results when these vary
(e.g. density and chemical inhomogeneities, matter-bounded geometry, 
distribution of dust and sources, atomic data, etc). 
However, although it is beyond the scope of this paper to discuss in detail the effect of these factors in the
derivation of the resulting abundances, in Section 4 we discuss the impact on our results of two of these parameters, as it is the electron density and the $\alpha_{ox}$ parameter  as these are two of the main driver factors that can affect the final results (e.g. \cite{dorsuv}). In the first case we also built a grid
with a higher electron density of 2\,000 cm$^{-3}$, and in the second with an $\alpha_{ox}$ = -1.2,
which is near the mean value found by \cite{miller11}.
In all case the code here presented admits new and modified grids that can allow us to study other sources of uncertainty in the models.

Overall, the models cover the range of 12+log(O/H) from 6.9 to 9.1 in bins of 0.1 dex.
In addition we consider values of log(N/O) from -2.0 to 0.0 in bins of 0.125 dex. Besides all models consider values of log $U$ from -4.0 to -0.5 in bins of 0.25 dex.
This implies to extend towards higher values of log $U$ the range originally defined in  \cite{hcm14}
 for star-forming regions as it is expected to find
very high values of this parameter in AGNs (e.g. \citealt{matsuoka18}).
This gives a total of 5\,865 models in the grid.
In addition to improve the model resolution as it is discussed for star-forming regions in \cite{pm16}, the code also allows to interpolate linearly the fluxes predicted by the
models by a factor 5
in the three running input variables in the grid (i.e. O/H, N/O, and log $U$). We also discuss in
Section 4 the impact of the enhancement of the resolution of the grid in the results.

In Figure \ref{bpt} we show two of the so-called BPT \citep{bpt} diagrams,
traditionally used to classify star-forming galaxies and AGNs, representing
the emission line ratios [\oiii]/\hb vs. [\nii]/\ha\, and [\oiii]/\hb vs. [\sii]/\ha\
for the control sample as compared with the results of the whole grid of models. 
As can be seen the objects lie in the right upper region of the diagrams over the separation curves defined by \cite{kauffman03} and \cite{kewley01}
and  most of this sample is well covered by the grid.
A large fraction of the models of the grid occupies the region usually assigned to star-forming objects, but this position is apparently controlled by the metallicity of
the models, as also discussed in \cite{kewley01}, so the observed position of Sy2 galaxies in these diagrams is possibly of empirical origin.

\begin{figure*}
\centering

\includegraphics[width=8cm,clip=]{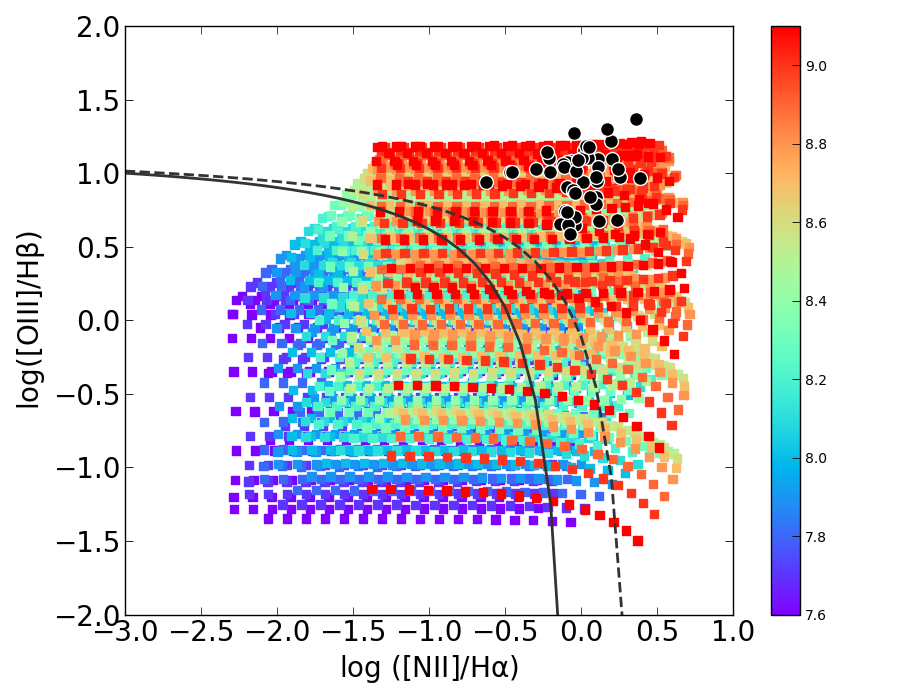}
\includegraphics[width=8cm,clip=]{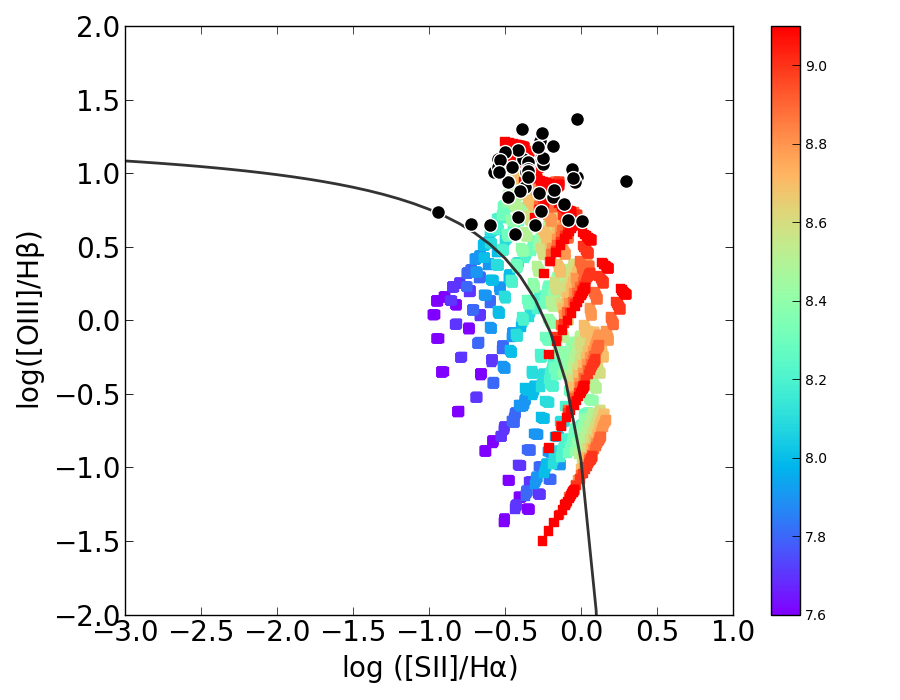}

\caption{Diagnostic diagrams showing the emission-line ratio
[\oiii]/\hb\ in relation to [\nii]/\ha\ (left) and to [\sii]/\ha\ (right) both for the
control sample (black circles) and the whole grid of models (colored squares). 
The color bar represents the metallicity of each model.
The solid line represents the curve defined
by Kauffman et al (2003) to separate AGNs and star-forming regions. The dashed red line represents the line defined by
Kewley et al (2001) to separate AGNs from pure star-forming and composite regions.}

\label{bpt}
\end{figure*}

\subsection{The {\em HCm} code adapted for AGNs}

The program {\sc HCm}\footnote{Publicly available in the
webpage \url{http://www.iaa.csic.es/~epm/HII-CHI-mistry-agn.html}} was done using {\sc python} to calculate O/H and N/O
chemical abundances ratios and log$U$ using a bayesian-like comparison between the predictions from the models described
above and a set of observable emission lines.
This code was originally designed to be used in star-forming galaxies but in this work we describe its application
and the specific features taken into the account for its use in the NLR of AGNs.
The availability of the code ensures its application for large samples of objects and its
reproducibility for different samples of objects. At same time it allows to change the library models adapting them to different initial assumptions for the calculation.

The code uses as observable inputs the reddening corrected relative-to-\hb\ emission lines
from [\oii] $\lambda$3727 \AA, [\neiii] $\lambda$3868 \AA,[\oiii] $\lambda$4363 \AA, [\oiii] $\lambda$5007  \AA, [\nii] $\lambda$6583 \AA, and
[\sii] $\lambda$$\lambda$6717+6731 \AA\ with their corresponding errors. However, the code also provides  a solution in case one or several of these
are not given, what implies that the assumptions made to perform the corresponding calculations and the final
derived uncertainty vary accordingly as it is described in the sections below.

\subsubsection{Derivation of N/O}

In a first iteration the code searches for N/O as a
weighted mean over all the space of models: since N/O can be estimated using
optical emission-lines of similar excitation, it can be calculated without any specific assumption about the ionization parameter and helps
to constrain the space of models to use [\nii] lines to derive oxygen abundance in a second later iteration.

\begin{equation}
\log({\rm N/O})_f = \frac{\sum_i \log({\rm N/O})_i/\chi^2_i}{\sum_i 1/ \chi^2_i}
\end{equation}

\noindent where log(N/O)$_f$ is the final derived nitrogen-to-oxygen ratio, log(N/O)$_i$ are the
values for each one of the individual models
and the $\chi$ values are assigned for each model as:

\begin{equation}
\chi_i =\sum_j \frac{(O_j - T_{ji})^2}{O_j}
\end{equation}

\noindent being the normalized quadratic difference between certain observed emission line ratios, $O_j$, and
the corresponding prediction from each model, $T_j$. 
The errors are calculated as the quadratic sum of the standard deviation of the $\chi$-weighted
resulting distribution and that from a Monte Carlo simulation carried out using the random deviation
in each line from the input emission line uncertainties.

In the case of N/O the code uses two different observable sensitive emission-line ratios. \cite{pmc09} suggested the use of N2O2, defined as:

\begin{equation}
{\rm N2O2} = \log \left( \frac{[{\rm N \textsc{ii}}] \lambda6583}{[{\rm O\textsc{II}}] \lambda3727} \right),
\end{equation}

\noindent and a similar version more useful for restricted observed spectral ranges, 
and hence less sensitive to reddening correction or flux calibration, the N2S2 parameter, defined as:

\begin{equation}
{\rm N2S2} = \log \left( \frac{ [{\rm N\textsc{ii}}] \lambda6583}{[{\rm S\textsc{ii}}] \lambda\lambda6717+6731} \right)
\end{equation}

These parameters have the advantage that they do not present almost any dependence on excitation as they
only involve low-excitation lines. Indeed they have been proposed for star-forming objects as direct
tracers of the metallicity, based on the assumption that secondary production of N at high-metallicity makes N/O to be a direct estimator of O/H.
N2O2  has been also proposed as estimator of total metallicity in NLR of Sy2 galaxies
by \cite{castro17} using results from photoionization models and an assumption about
the expected relation between oxygen and nitrogen relative abundances.

In Figure \ref{n2-no} we see the behavior of these two parameters with N/O both from some of the models of the grid and from the estimates
obtained  by \cite{dors17}  for the control sample.
 As it can be seen the models do not show a large
dispersion as a function of metallicity or excitation and the objects tend to adopt a linear relation in both cases with N/O.
In addition we see that the adopted relations for star-forming objects by \cite{pmc09} are not valid in the case of NLRs of AGNs and
new relations can be adopted for this kind of objects.

In this way, as a sub-product of the models,  a linear fitting to them 
can be used to derive in an alternative  direct way the N/O ratio, yielding} on one hand, for N2O2:

\begin{equation}
\log ({\rm N/O}) = (0.97 \pm 0.01) \cdot {\rm N2O2} - (0.50 \pm 0.01) 
\end{equation}

\noindent which is also represented in Figure \ref{n2-no}. The mean offset of the objects analyzed in \cite{dors17}
is only of 0.07 dex higher  with a standard
deviation in the residuals of 0.13 dex.

Regarding N2S2, the linear fitting to the whole grid of models gives:

\begin{equation}
\log({\rm N/O}) = (0.88 \pm 0.01) \cdot {\rm N2S2} - (0.69 \pm 0.01) 
\end{equation}

In this case, the mean offset of the N/O values derived using this expression in relation to values derived by
\cite{dors17} is 0.05 dex higher with
a standard deviation of the residuals of 0.12 dex.

\begin{figure*}
\centering

\includegraphics[width=8cm,clip=]{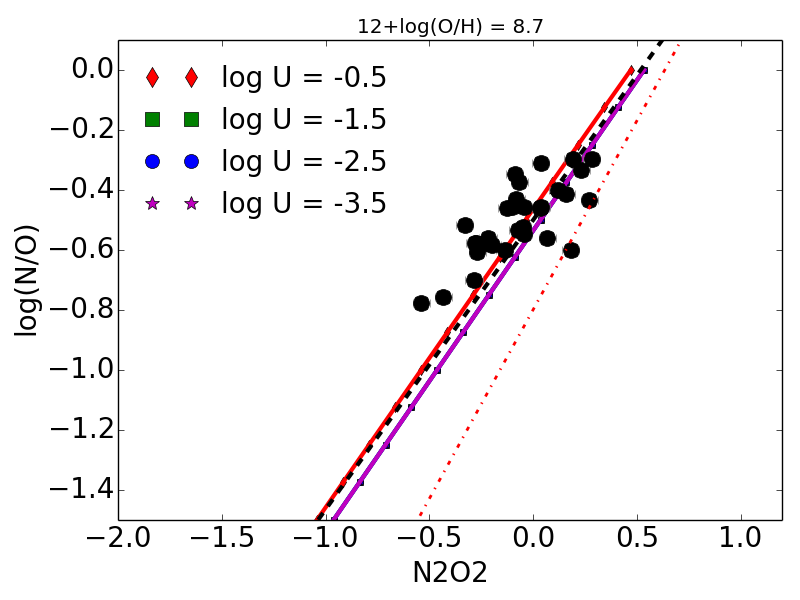}
\includegraphics[width=8cm,clip=]{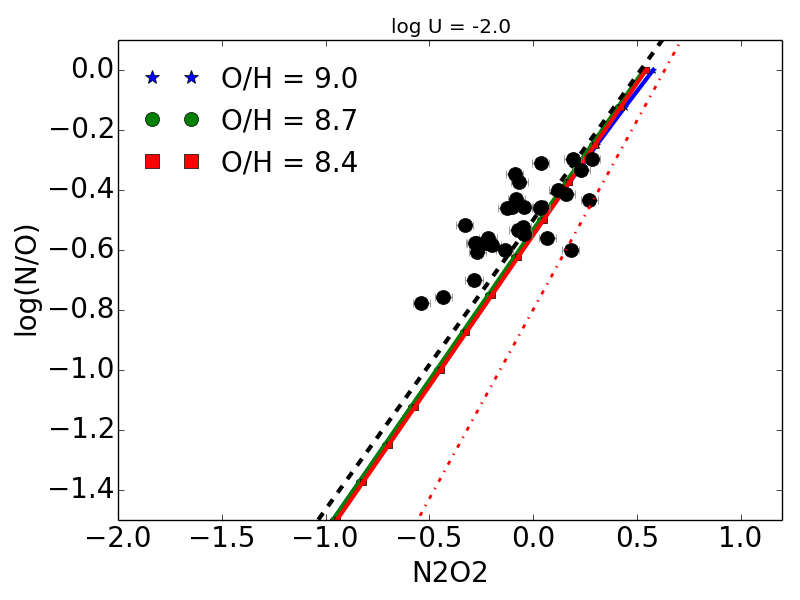}
\includegraphics[width=8cm,clip=]{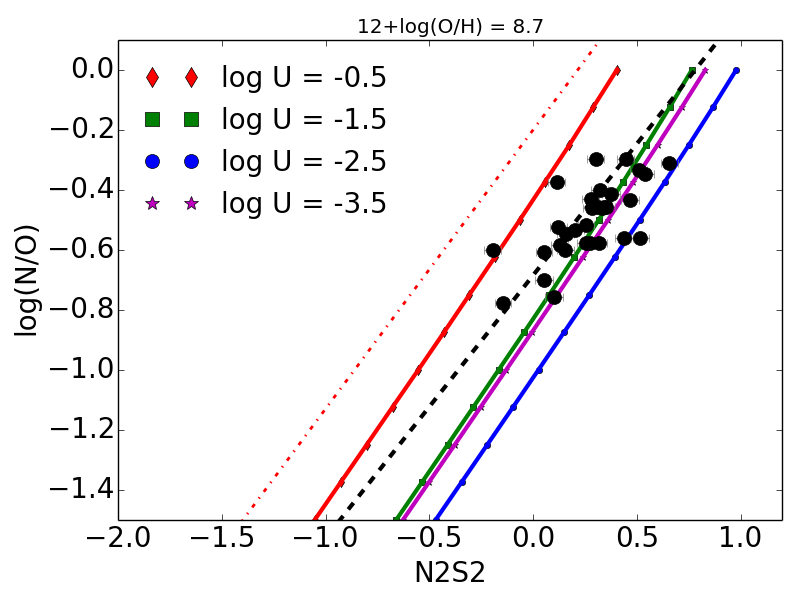}
\includegraphics[width=8cm,clip=]{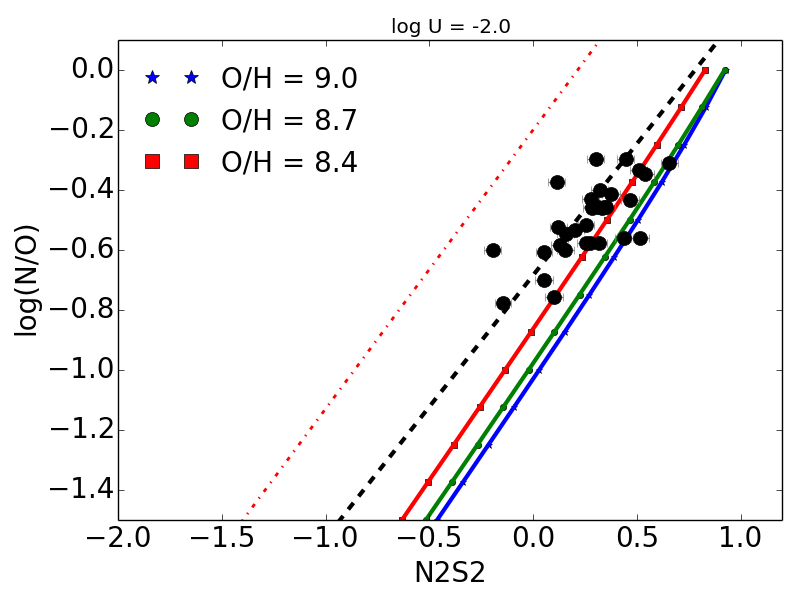}

\caption{Relation between the elemental abundance ratio N/O with the emission line ratios N2O2 (upper panels) and N2S2 (lower panels)
both for some of the models of the grid and for the control sample with the abundances derived by
 Dors et al (2017). The panels in the left column show results from the models
described in the text at a fixed O/H = 8.7 and the panels at right column at a fixed log($U$) = -2.0.
The dashed black lines in all panels represent the linear fitting to the models
while thered point-dashed lines represent the empirical  calibrations of these 
relations for star-forming objects in P\'erez-Montero \& Contini (2009).}

\label{n2-no}
\end{figure*}

\subsubsection{Derivation of O/H and U}

Once N/O has been estimated, the code begins a new iteration through the space of models constrained to the
N/O values and the derived uncertainty. The code now calculates the values of the
total oxygen abundance and the ionization parameter as the weighted means over all
the considered models as:

\begin{equation}
12+\log({\rm O/H})_f = \frac{\sum_i (12+\log({\rm O/H}))_i/\chi^2_i}{\sum_i 1/ \chi^2_i}
\end{equation}

\noindent and

\begin{equation}
\log(U)_f = \frac{\sum_i \log(U)_i/\chi^2_i}{\sum_i 1/ \chi^2_i}
\end{equation}

\noindent where the (O/H)$_f$ and $U_f$ are the resulting values, and the
(O/H)$_i$ and $U_i$ are the individual values for each model.
The $\chi^2$ values and the corresponding errors are calculated in the same way as it was 
explained for N/O. We now describe the observables used for the calculation of the
$\chi_i$ values, including:

\begin{equation}
{\rm RO3} = \frac{[{\rm O\textsc{iii}}] \lambda5007 \AA}{[{\rm O\textsc{iii}}] \lambda4363 \AA}
\end{equation}

\noindent what has a direct relation with the electron temperature of the gas. In NLRs of AGNs
the direct relation of this ratio with total oxygen abundance
is not well established but, according to models, can be also used as a proxy for the presence of
heavy elements in the gas.  The main problem in this scenario is that, contrary to star-forming regions,
models predict that anon-negligible fraction of the oxygen total abundance correspond to higher ionization stages whose emission lines
are not detectable in the optical range. Therefore it is not evident to
derive any direct relation between the relative intensities of the strongest optical emission-lines and the total abundances of the elements emitting them.

\begin{figure}

\centering

\includegraphics[width=8cm,clip=]{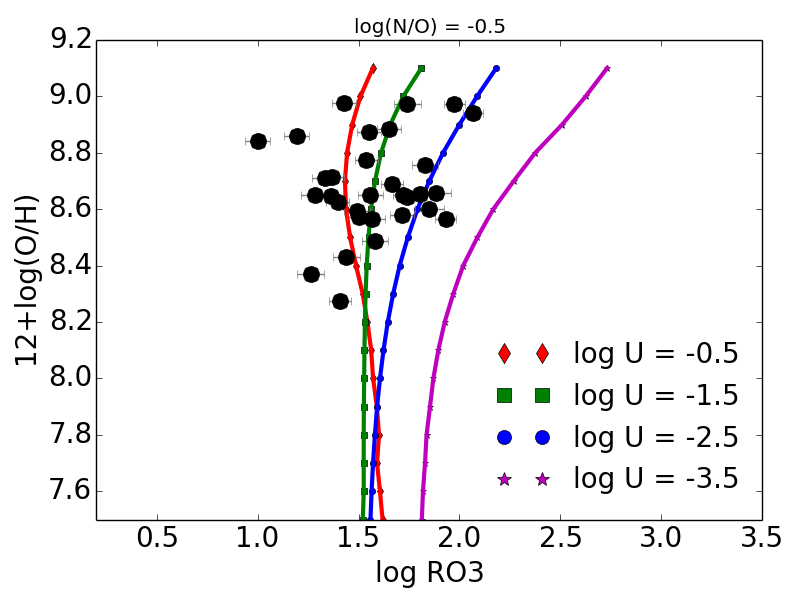}

\caption{Relation between the total oxygen abundance and
the logarythm of  the emission line ratio
[\oiii] $\lambda$$\lambda$5007/4363 \AA\ both for the abundances derived by   Dors et al (2017) for the control sample, as represented with black circles, and for models,
for different values of $U$, at a fixed log(N/O) value of -0.5.}

\label{ro3-oh}
\end{figure}

In Figure \ref{ro3-oh} we show the relation between 
the logaritm of this emission-line ratio
and the abundances derived by \cite{dors17} for the control sample along with some of the models of the grid
at a fixed log(N/O) = -0.5.. The models cover the values estimated in these objects, but the direct relation with 
metallicity is only clear at high values, when the
 ions of the metals whose lines are observed in the optical range begin to act as effective coolants of the gas and they have
a clear influence on its electron temperature, what in star-forming regions occurs
at much more lower values of O/H.

In relation to other observables used by the code based on nitrogen lines, since N/O has been fixed in the space of models
by the code in the previous step described above,
the [\nii] emission lines can be now used to derive oxygen abundances.
Otherwise, if the studied object lies out of the expected chemical relation 
between O/H and N/O we can obtain wrong derivations of O/H.

Among the estimators of O/H based on [\nii] used as observables by the code are the N2 parameter  \citep{storchi,dtt02}, defined as:

\begin{equation}
{\rm N2} = \log \left( \frac{[{\rm N\textsc{ii}}] \lambda6583}{H\alpha} \right)
\end{equation}

This ratio has the advantage that does not depend on reddening correction and it is easy to measure in the red part of
the spectrum with an adequate resolution. This emission-line ratio has been also
pointed out as a good tracer of total metallicity in NLRs of AGNs by \cite{storchi98}, who also point out its relatively lower 
dependence on ionization parameter. However its use as a tracer of O/H relies on the assumption of a direct expected relation between O/H and N/H.

\begin{figure*}
\centering

\includegraphics[width=8cm,clip=]{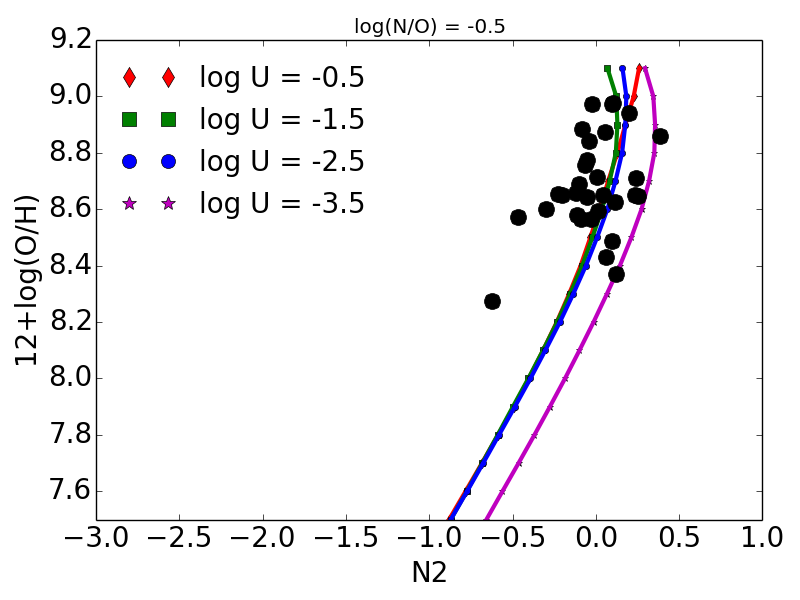}
\includegraphics[width=8cm,clip=]{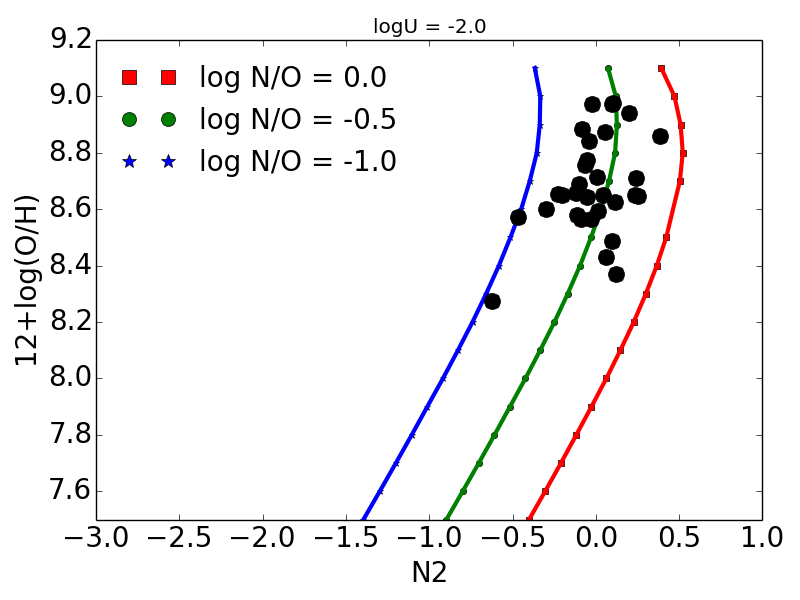}
\includegraphics[width=8cm,clip=]{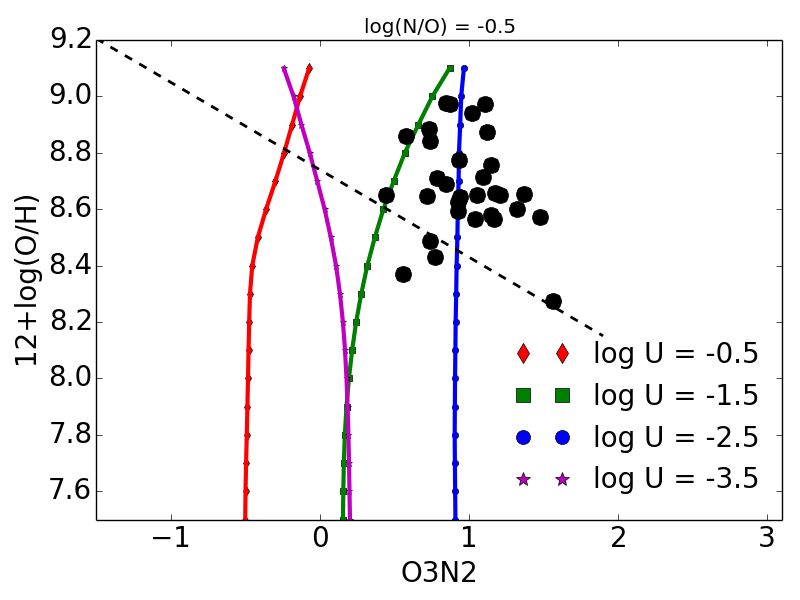}
\includegraphics[width=8cm,clip=]{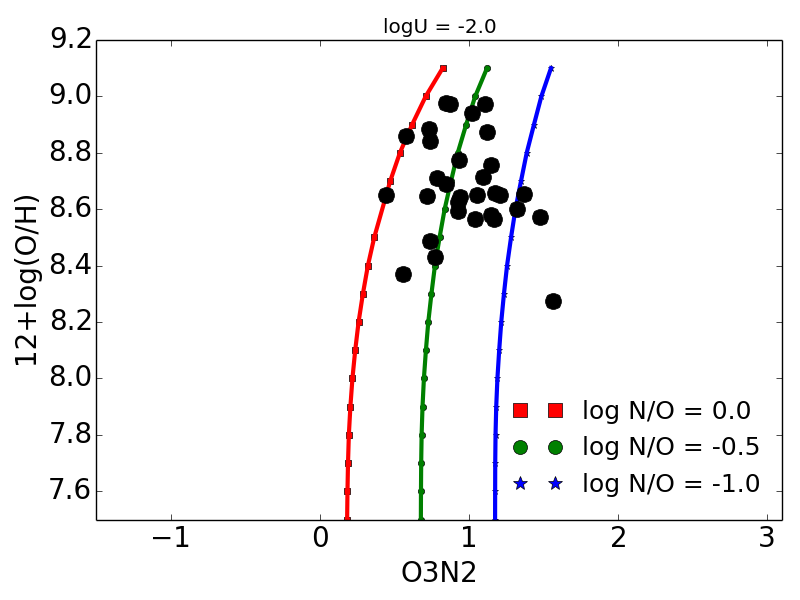}

\caption{Relation between the total oxygen abundance 12+log(O/H) with the emission line ratio N2 (upper panels) and O3N2 (lower panels)
both  for the abundances derived by Dors et al. (2017) for the control sample
and from some of  the  models
described in the text at a fixed log(N/O) = -0.5 (left panels) and  at a fixed log($U$ )= -2.0 (right panels).
The dashed black lines represent the empirical calibrations for star-forming regions derived by P\' erez-Montero \& Contini (2009).}

\label{n2-oh}
\end{figure*}

In Figure \ref{n2-oh} we see the relation between O/H and N2 both for the abundances derived by \cite{dors17} for the control sample and for the predictions made
by some of the models of our grid for different values of $U$ and N/O. The relation is monotonically growing up to very high values of metallicty but it presents a large dependence 
on N/O, as in the case of star-forming regions as pointed out by \cite{pmc09}. The dependence on $U$ is slightly lower but can also be important.
We also see that the empirical calibration given for star-forming regions cannot be used for this kind of objects, as it underestimates the
derived metallicity.

Other observable based on [\nii] emission lines that we can use to constrain both O/H and $U$ is O3N2, defined by \cite{alloin79} as:

\begin{equation}
{\rm O3N2} = \log \left(\frac{[{\rm O\textsc{iii}}] \lambda5007}{H\beta} \cdot \frac{H\alpha}{[{\rm N\textsc{ii}}] \lambda6583} \right)
\end{equation}

The relation of this parameter with total metallicity  is also shown in Figure \ref{n2-oh} both for the studied sample of galaxies and for models attending to their
dependence on N/O and $U$. As can be seen it strongly depends on both parameters given the
large correlation of the [\oiii]/\hb\ ratio with excitation \citep{storchi98}.
On the other hand, the relation of this parameter with O/H is totally different to the behavior observed in star-forming
regions where, for the high-metallicity regime, the parameter decreases for larger values of O/H and remains relatively unaffected for
low values of $Z$, so the linear relation is not usually defined for high values of O3N2 (i.e. O3N2 $>$ 2). In contrast, for NLRs of AGNs this parameter grows with O/H above all for large metallicity and high $U$ because the relative abundance of the ions
emitting in the optical part of the spectrum begins to be important and play a more relevant role in the cooling of the gas at this regime.

In addition, among the observables used by the code to derive O/H and log($U$
are the R23 parameter, defined by \cite{pagel79} as:

\begin{equation}
{\rm R23} = \frac{[{\rm O\textsc{ii}}] \lambda 3727 + [{\rm O\textsc{iii}}] \lambda\lambda 4959 + 5007}{H\beta}
\end{equation}

\begin{figure*}
\centering

\includegraphics[width=8cm,clip=]{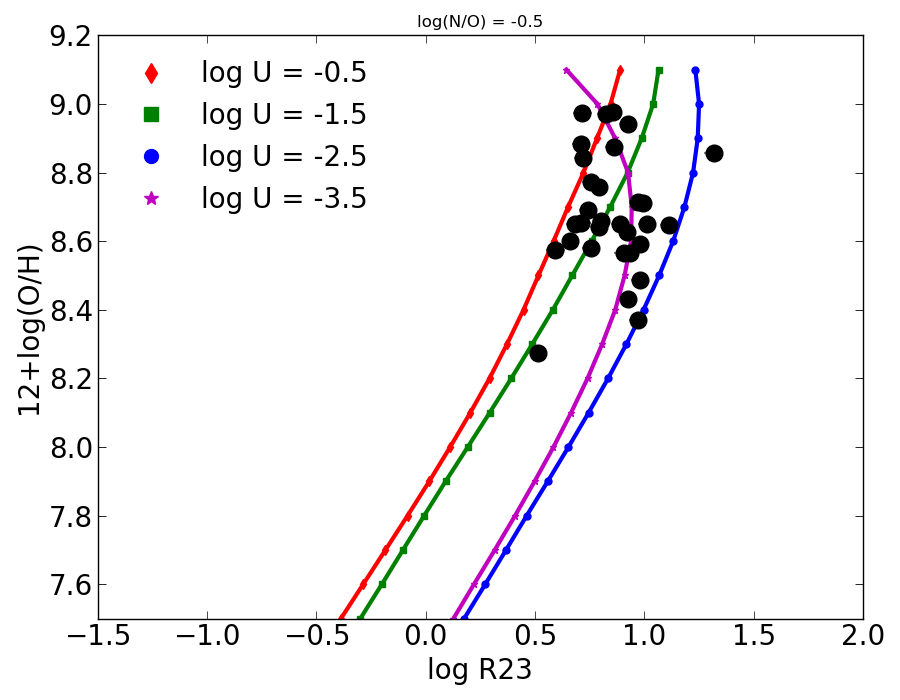}
\includegraphics[width=8cm,clip=]{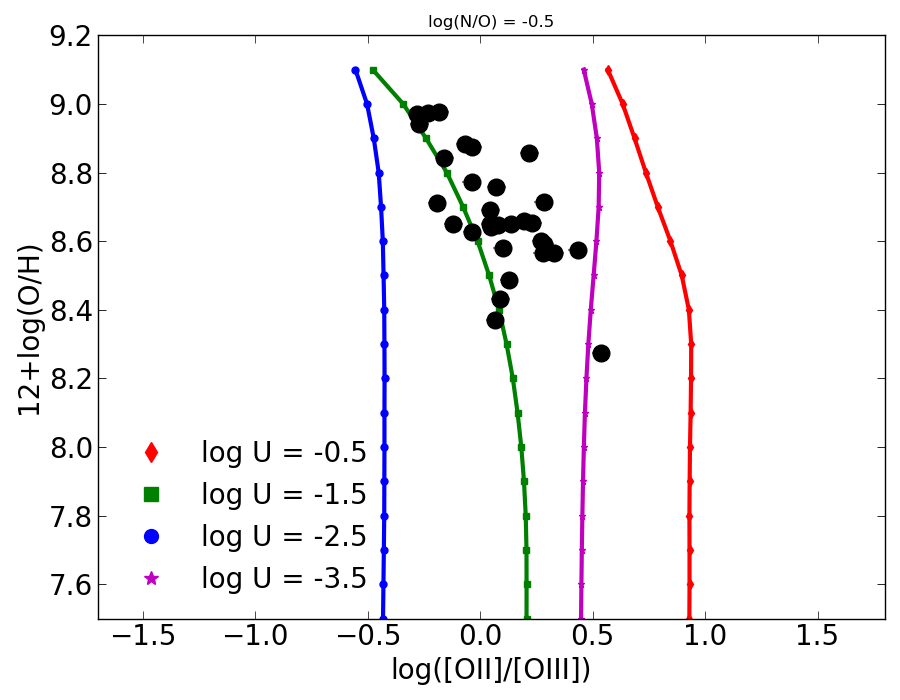}
\includegraphics[width=8cm,clip=]{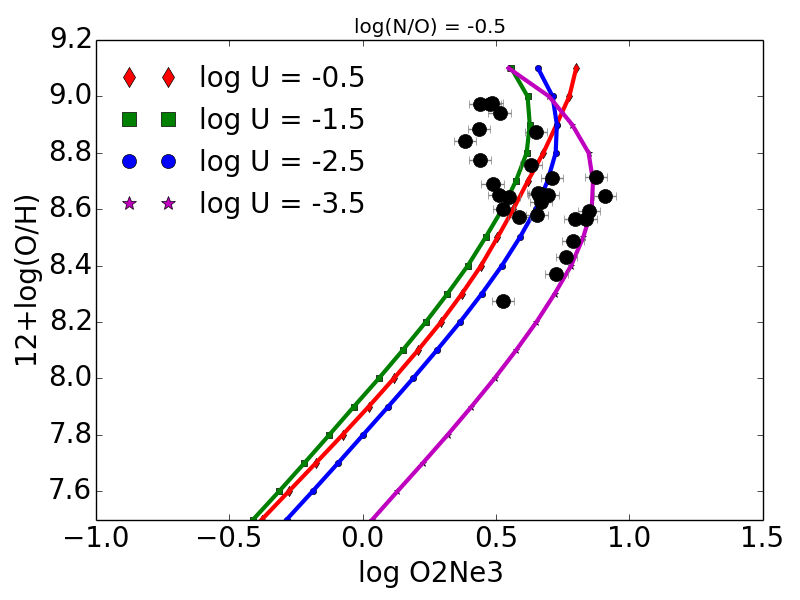}
\includegraphics[width=8cm,clip=]{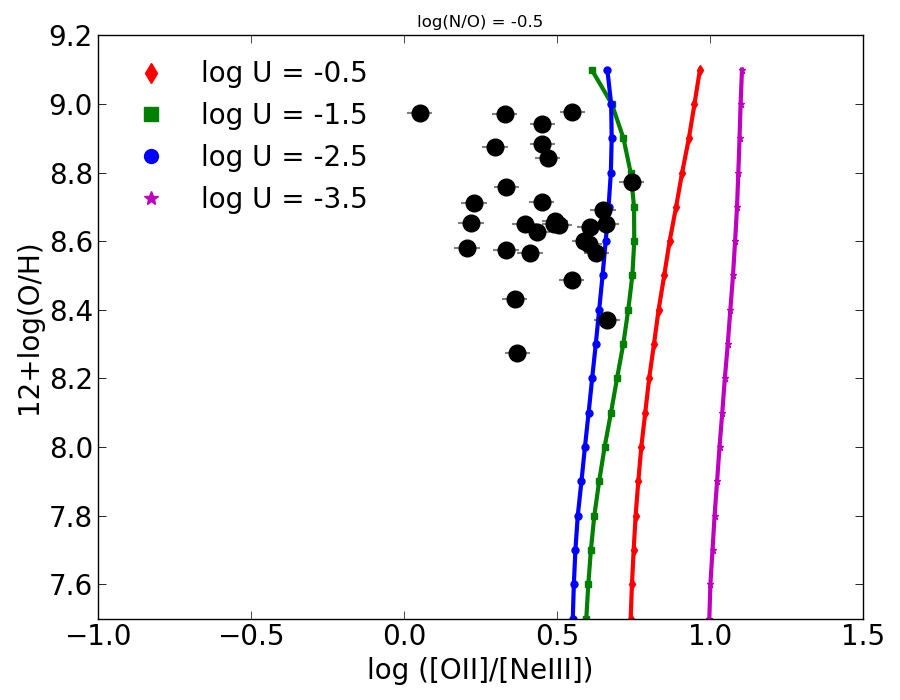}

\caption{Relation between the total oxygen abundance 12+log(O/H) as derived in Dors et al (2017) for the control sample
with the emission line ratios R23 (left upper panel),
[\oii]/[\oiii] (right upper panel), O2Ne3 (left lower panel), and [\oii]/[\neiii] (right lower panel).
All panels also show results from some of the models
described in the text at a fixed log(N/O) = -0.5. .}

\label{oh-r23}
\end{figure*}

Note that this parameter is traditionally defined using the other strong [\oiii] line at $\lambda$4959 \AA,
so in the case of our code, which uses the theoretical ratio of this line  with $\lambda$5007 \AA\ (i.e. I(5007)/I(4959) = 3), this does not imply any difference if the ratio is calculated in 
the same way for the observable. The relation between log R23 for the sample of objects and the models are shown in Figure \ref{oh-r23}.
This parameter presents a totally different behavior for high values of $U$ to that observed in star-forming regions as it is not double-valued
(i.e. R23 increases for low O/H and decreases for high O/H). On the contrary, it  presents a
monotonically growing relation with O/H up to very high values.
As previously explained for other observables based on [\oiii] lines this is mostly due to the fact that, according to our models, most of oxygen keeps in higher ionization stages than those whose
emission-lines can be measured in the optical part of the spectrum. As a consequence, these lines act as effective coolants of the gas
in AGNs at a much higher metallicity than in the case of star-forming regions. In addition it is observed, as in the case of star-forming regions (e.g. \citealt{pmd05}) that this ratio has a strong
dependence on $U$, what can even affect the shape and the turnover of this relation.

To minimize this dependence, the code also considers the ratio [\oii] $\lambda$3727/[\oiii] $\lambda$$\lambda$4959+5007, what
helps to restrict the values of the ionization parameter. This is also shown in Figure \ref{oh-r23},
where we can see that this ratio presents  a dependence on metallicity much lower than in the case of R23.

\begin{figure*}
\centering

\includegraphics[width=8cm,clip=]{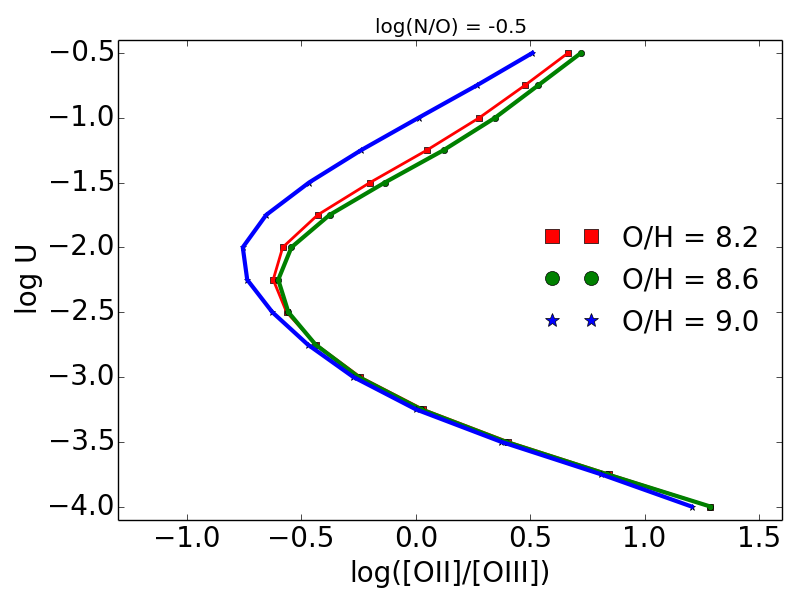}
\includegraphics[width=8cm,clip=]{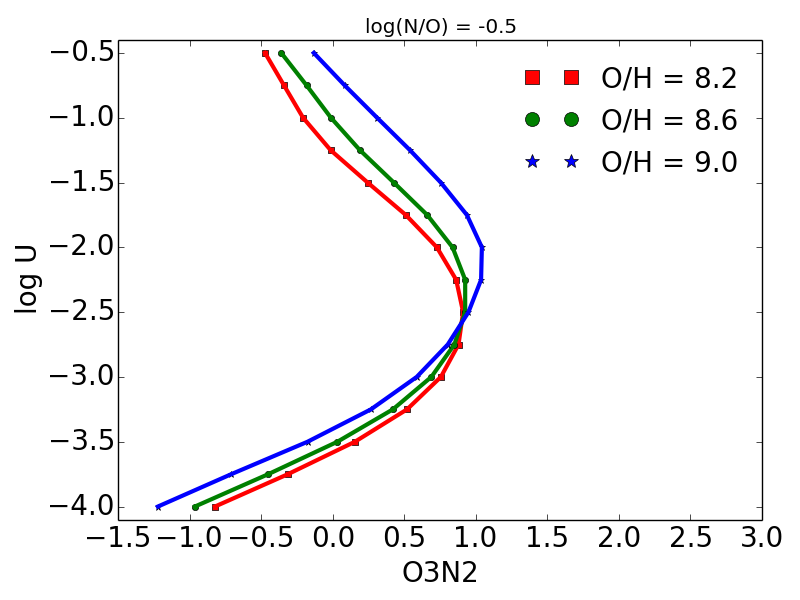}

\caption{Relation between the ionization parameter and the emission-line ratios [\oii]/[\oiii] (at left), and O3N2 (at right) as predicted
from photoionization models for different metallicities and assuming a constant N/O value at -1.0.}

\label{rel-U}
\end{figure*}

Nevertheless the dependence of this emission-line ratio on $U$ is not monotonic as can be seen in Figure \ref{rel-U} as predicted from models. 
The ratio decreases as $U$ increases, as it is expected and it occurs in star-forming regions, but, for values of log($U$) larger than -2.0, it begins to
turn and it increases giving place to a double-valued relation. This bivaluated behavior is also seen for other observables based on [\oiii] lines, as it is the case for O3N2 and can also be seen
in the same Figure. This simply happens because at high values of $U$ the relative intensity of [\oiii] decreases rapidly as oxygen is ionized on higher stages.
Therefore to overcome the problem of a degeneration in the derivation of $U$ using this emission-line ratio we do not consider in our models
all values with log($U$) lower than -2.5, that is the value for which O3N2 reaches it maximum.
The range of log $U$ considered in our models cover the typical values
in other works studying similar objects (e.g. \citealt{dopita14}),
and does not affect the determination of N/O and O/H as shown in the next section.

Finally, the code also allows to use the [\neiii] $\lambda$3868 \AA\ emission line, but it is only used
as a alternative to [\oiii] $\lambda$5007 \AA\ when this is not given (i.e. for specific setups where [\oiii] is not covered), as these two ions present a
quite similar ionization structure and the [\neiii] line can be used as a proxy for
the intensity in the high-excitation region at very blue wavelengths, as discussed in \cite{pm07}.
In this way, the O2Ne3 can be defined as:

\begin{equation}
{\rm O2Ne3} = \frac{[{\rm O\textsc{ii}}] \lambda 3727 + [{\rm Ne\textsc{iii}}] \lambda3868}{H\beta}
\end{equation}

Note that in this definition, contrary to \cite{pm07}, no empirical factor has been considered for
the intensity of [\neiii], as this is not required by the code to compare the observed fluxes with the results from models.
The relation between the O2Ne3 parameter with O/H, and the same relation with the
 [\oii]/[\neiii] as an indicator of the excitation are also shown in the lower panels of Figure \ref{oh-r23}.
As can be seen both observables behave in very similar way to those based on [\oiii] and therefore can be used instead
by the code to derive abundances and excitation in NLRs of AGNs.

\section{Results and discussion}

\subsection{Resulting properties of the sample}

\begin{figure*}
\centering

\includegraphics[width=8cm,clip=]{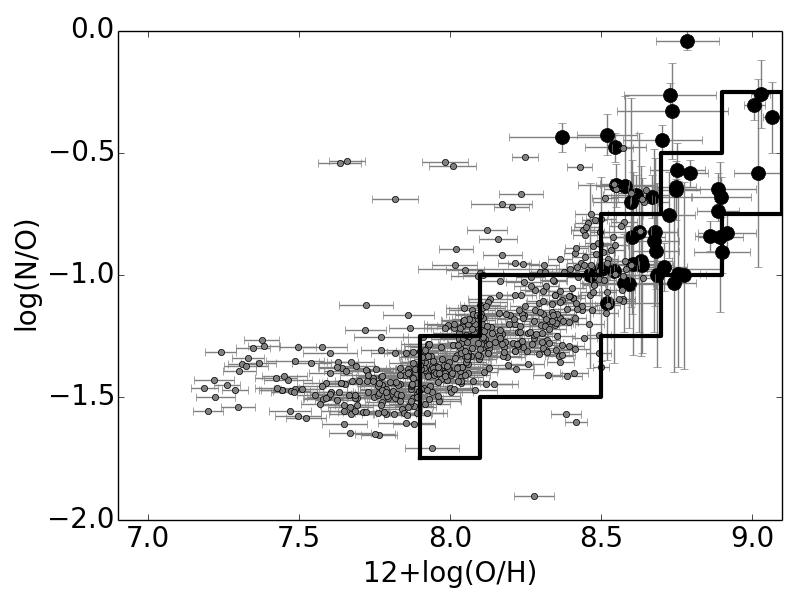}
\includegraphics[width=8cm,clip=]{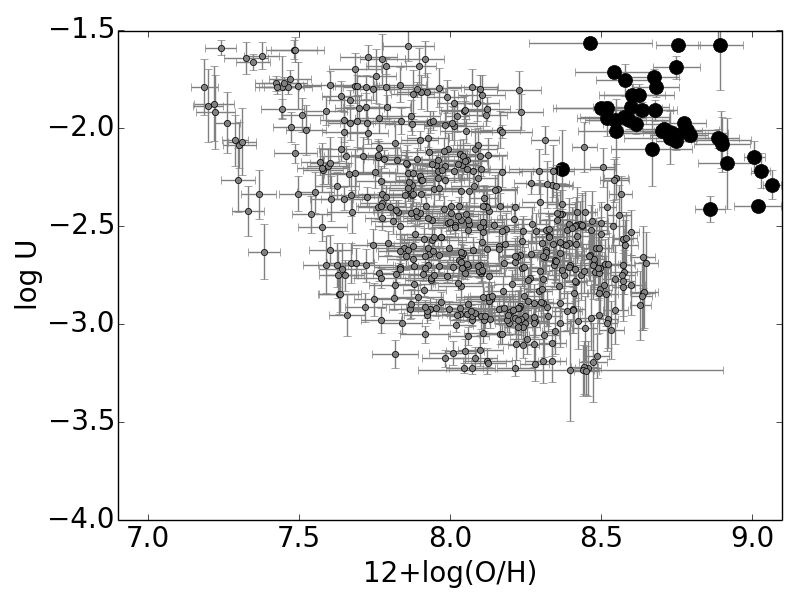}

\caption{Relations between O/H and N/O (left) and $U$ (right) for the resulting values
of our method for the control sample.
Grey points represent the sample of star-forming regions analyzed following
a similar bayesian-like procedure by P\'erez-Montero (2014).
The solid black line encompasses the space of models empirically constrained by P\' erez-Montero (2014) used to derive O/H when no previous derivation of N/O is performed.}

\label{oh-no}
\end{figure*}

In this subsection we discuss the results obtained by using
the {\sc HCm} code for the 47 Seyfert 2 and 1.9 galaxies presented in \cite{dors17})
when all the available emission-lines required by our code
are used.
The results of the code include too associated uncertainties which are quadratical additions of those from the standard deviation ot the $\chi^2$-weighted distribution
of the input grid values and the dispersion from the distribution of results after 100 Monte-Carlo iterations
using the input reddening-corrected  lines randomly
perturbed with their errors.

In left panel of Figure \ref{oh-no} we show the resulting O/H and N/O abundance ratios and their 
corresponding errors. The ranges of metallicity are quite similar to those obtained using the tailor-made models
presented in \cite{dors17},
going from 12+log(O/H) = 8.37 to 9.07 (from 0.5 to 2.4 times
the solar metallicity, taking as reference the value 12+log(O/H)$_{\odot}$ = 8.69 in \citealt{asplund09}). Regarding nitrogen, the range in log(N/O) goes from -1.11 to -0.04
(equivalent to the range 0.6 to 6.6 times the solar N/O ratio)
(taking as log(N/O)$_{\odot}$ = -0.82 from \citealt{asplund09}).
These values are in concordance with the expected relation between O/H and N/O in a regime of
secondary nitrogen production, giving place to an enhancing N/O ratio for higher metallicities (e.g. \citealt{henry2000}), although
with a large dispersion.
In the same Figure we compare the obtained abundances with those derived using {\sc HCm} for star-forming
objects \cite{hcm14} for the sample presented by \cite{marino13}. As can be seen the type 2 AGNs occupy the regions of high-metallicity
star-forming regions in agreement with the results obtained by  \cite{dors17}, but a non-negligible 
fraction of the analyzed objects lie in a region outside the expected relation what justifies the use of a previous determination of N/O to derive O/H abundances using [\nii] emission lines, instead of assuming any expected
relation between O/H and N/O.

Regarding the relation between metallicity and ionization parameter, in right panel of Figure \ref{oh-no} we
show the values obtained by our code for the same sample. In this case, the range of log$U$ looks to be much more constrained
(log$U$ goes from -2.42 to -1.27). 
In contrast to the relation between abundance ratios, there is not any
similarity with the behavior observed for star-forming regions as can be seen in the same plot. On one hand the average $U$ is much higher in the
case of AGNs than for the star-forming regions, even taking into account that the average metallicity of these is much lower. This behavior is also observed even if the grid of models is not constrained
to the higher values of $U$. The existence of a possible relation between O/H and $U$ as for star-forming objects, in the sense that
galaxies with higher metal content are in average less excited, is very uncertain and must be more deepley studied. In any case, unlike star-forming objects, the code
does not require any restriction in the space of models between O/H and $U$ when no [\oiii] 4363 \AA is introducied as input, as it will be discussed below.

\subsection{Comparison with the control abundances}

\begin{figure*}
\centering

\includegraphics[width=8cm,clip=]{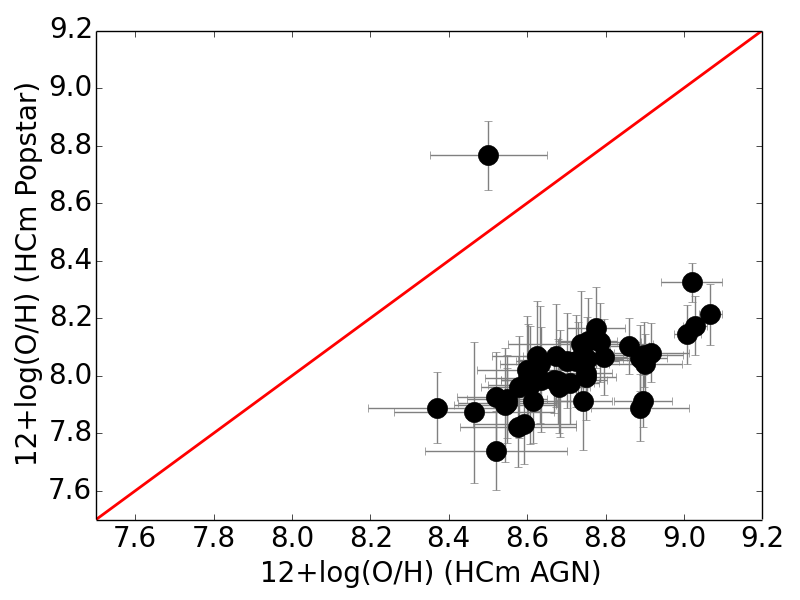}
\includegraphics[width=8cm,clip=]{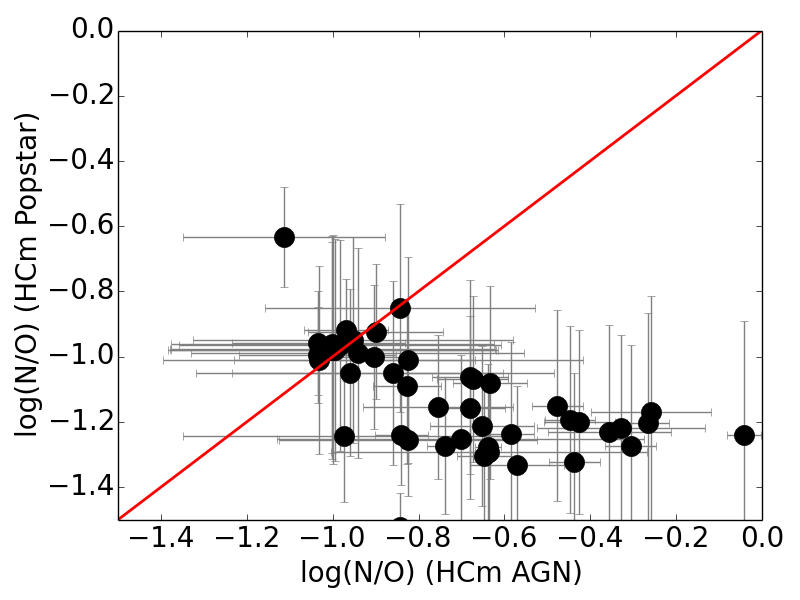}

\caption{Comparison between chemical abundances derived using the method described in this work and those
obtained assuming the ionizing spectral energy distribution of a massive young star cluster for the 
control sample. At left comparison of derived 12+log(O/H) and at right for log(N/O).
The red solid line represents the 1:1 relation.}

\label{comp_popstar}
\end{figure*}

\begin{figure*}
\centering

\includegraphics[width=8cm,clip=]{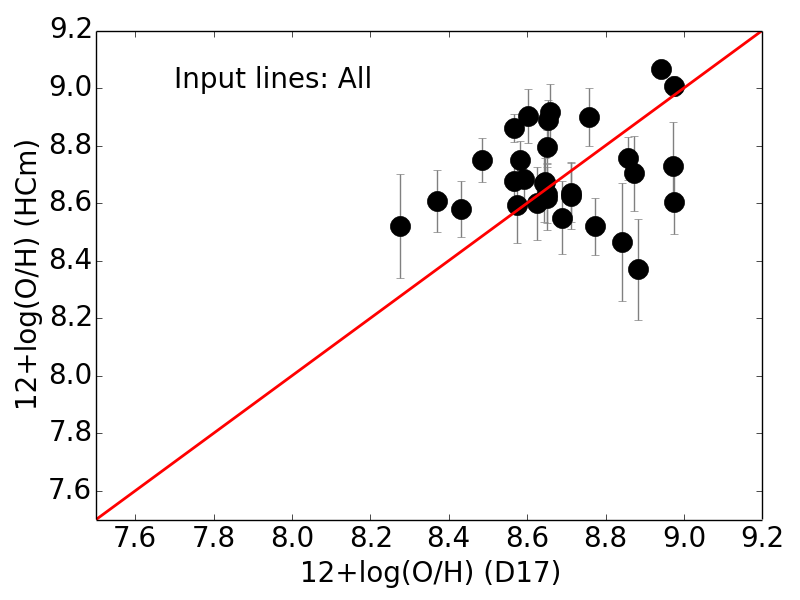}
\includegraphics[width=8cm,clip=]{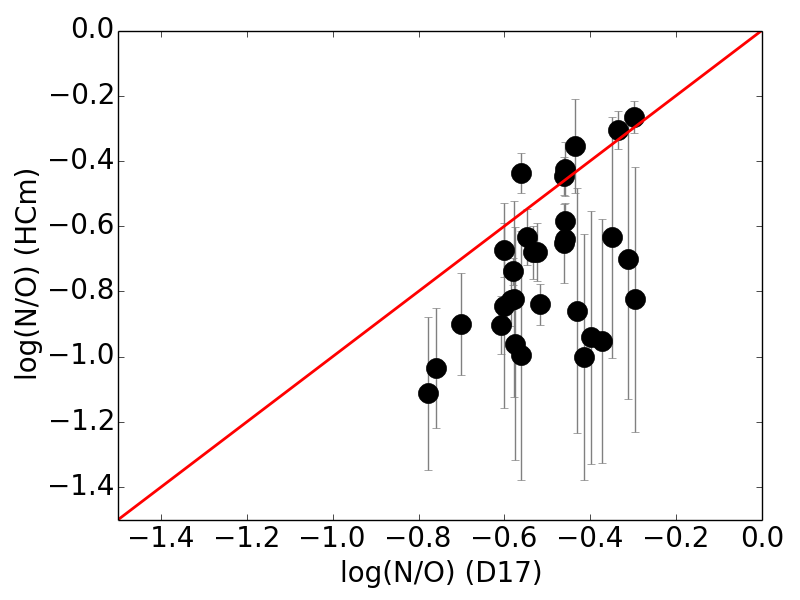}
\includegraphics[width=8cm,clip=]{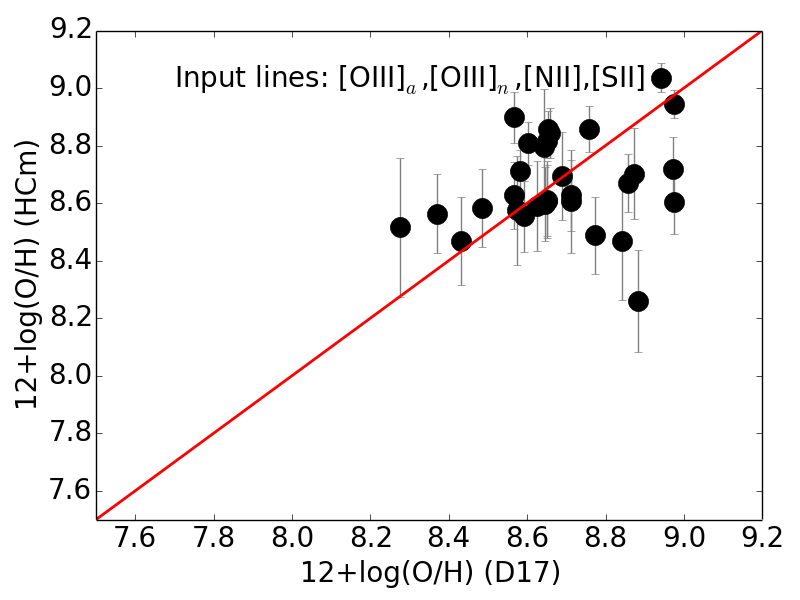}
\includegraphics[width=8cm,clip=]{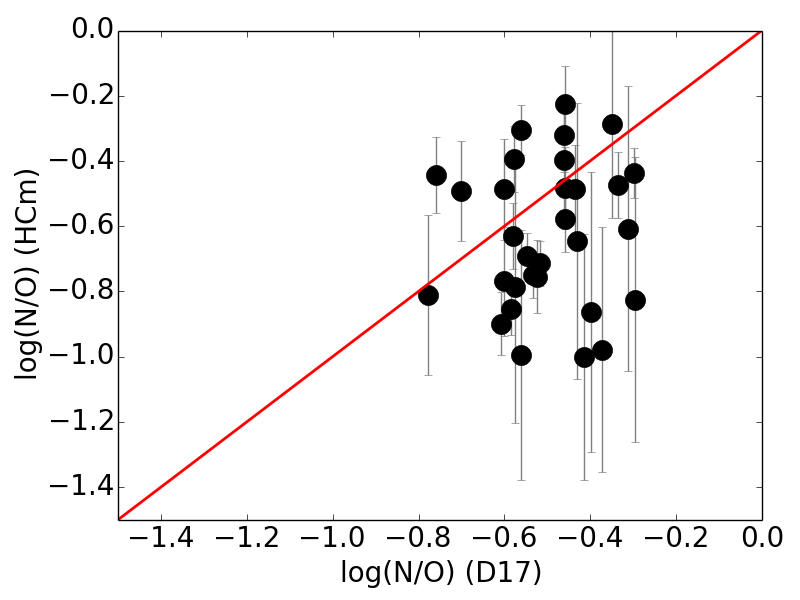}

\caption{Comparison between chemical abundances derived using the method described in this work and those
taken from Dors et al. (2017) from tailored photoionization models.
At left comparison of derived 12+log(O/H) and at right for log(N/O).
The upper panels show the comparison when all the lines are used while lower panels show the comparison in absence
of [\oii] $\lambda$3727 \AA.
The red solid line represents the 1:1 relation.}

\label{comp_dors}
\end{figure*}

In this subsection we discuss the deviations in the resulting chemical abundances obtained from {\sc HCm} code using
different sets of emission-lines taking as reference the abundances of the control sample obtained by a \cite{dors17}
using detailed tailor-made models.

The first comparison that we can perform as a test is to check to what extent the use of a specific SED for our
models impact on the determination of the final chemical abundances. To this aim we introduced the
compiled emission lines in the analyzed sample of Sy2 galaxies using the same code {\sc HCm}, but making
use of a SED of massive young star clusters taken from {\sc popstar} \citep{popstar}
with the features described in \cite{hcm14}. This comparison is shown
both for O/H and N/O in Figure \ref{comp_popstar}.
As can be seen the obtained
abundances when we assume a massive cluster SED is much lower both for O/H 
(0.7 dex in average) and N/O (0.4 dex in average) than the abundances derived from the same code assuming
the appropriate power-law SED. This comparison illustrates the importance of the input SED in the final results.

Continuing with our analysis of the results, In Figure \ref{comp_dors} we show the total oxygen abundances and the
nitrogen-to-oxygen ratio obtained from our code
using as input the appropriate AGN SED  as compared with the values derived by
\cite{dors17} and introducing all possible emission lines. In this way there is now a good  agreement
between the two sets  with average deviations
lower in most cases than the average uncertainty and with dispersions of the same order,
although it is noticeably worse for N/O.
Both the mean and the standard deviation of the residuals for the analyzed Sy2 galaxies are listed in Table \ref{dispersion}.
From this table it is easy to establish a comparison between the results from our
code using all emission lines and only different subsets.

In the lower panels of the same Figure we show the comparisons when the emission-line of [\oii] at $\lambda$3727 \AA\ is not
included as input in the {\sc HCm} code. This is usual in some setups when the very blue
part of the
spectrum is not available (e.g. in the Sloan Digital Sky Survey at very low redshifts).
As can be seen the agreement in this case is good, with deviations from
the abundances in the control sample lower in all cases than the usual obtained uncertainties.

\begin{table*}
\begin{minipage}{180mm}
\begin{center}
\caption{Mean and standard deviation of the residuals of the comparison between the resulting O/H and N/O
abundance ratios derived for {\sc HCm} using all involved lines and using only a constrained set of them
as compared with the values for the same objects obtained in Dors et al (2017).
In the table [\oii] stands for $\lambda$3727 \AA, [\neiii] for $\lambda$3868 \AA, [\oiii]$_a$ for $\lambda$4363 \AA, [\oiii]$_n$ for $\lambda$5007 \AA,
[\nii]  for $\lambda$6583 \AA, and [\sii] for $\lambda$$\lambda$6717+6731 \AA\AA.}

\begin{tabular}{clcccc}
\hline
\hline
&  Set of input emission lines  &  Mean $\Delta$(O/H) & St.dev. $\Delta$(O/H) &  Mean $\Delta$(N/O) & St.dev. $\Delta$(N/O) \\
\hline
  &  All lines &  - 0.01  &  0.21  &   + 0.23  &  0.19 \\ 
&  [\oiii]$_a$, [\oiii]$_n$, [\nii], [\sii] &  + 0.02  &  0.21&   + 0.12 &   0.24\\
&  [\oii], [\oiii]$_n$, [\nii], [\sii] &  - 0.08 &  0.32 &   + 0.08 &   0.12 \\
& [\oiii]$_n$, [\nii], [\sii]  &  +0.15  &  0.26  &  - 0.11  &  0.11  \\
& [\nii], [\sii]  &  + 0.29   &   0.29     &  - 0.12 &   0.10  \\
& [\oiii]$_n$, [\nii]  &  -0.24  &  0.15  &  - --  &  --  \\
& [\nii]  &  - 0.25   &   0.16     &  - -- &   --  \\
& [\oii], [\oiii]$_n$  &  - 0.11  &  0.21   &   --    &  --  \\
& [\oii], [\neiii]   &   - 0.13   &   0.30   &   --   &  --   \\
\hline

\label{dispersion}
\end{tabular}
\end{center}
\end{minipage}
\end{table*}

\begin{figure*}
\centering

\includegraphics[width=8cm,clip=]{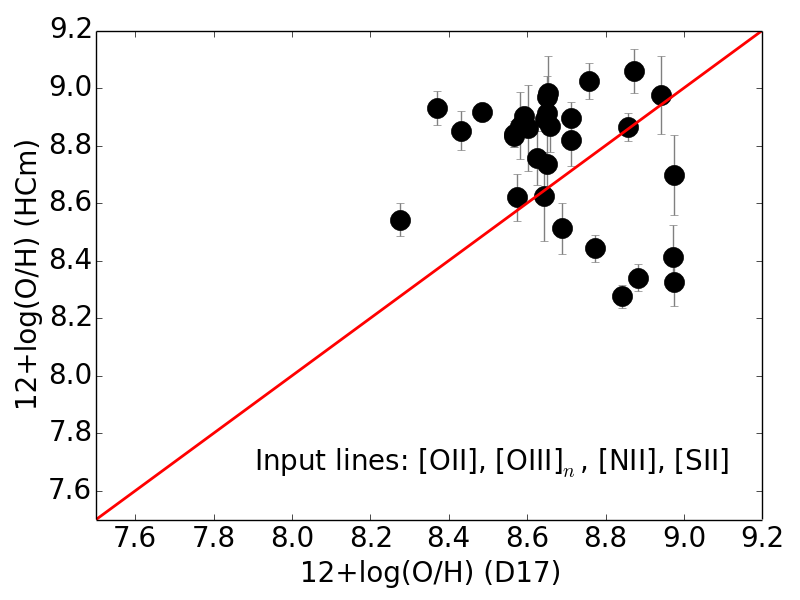}
\includegraphics[width=8cm,clip=]{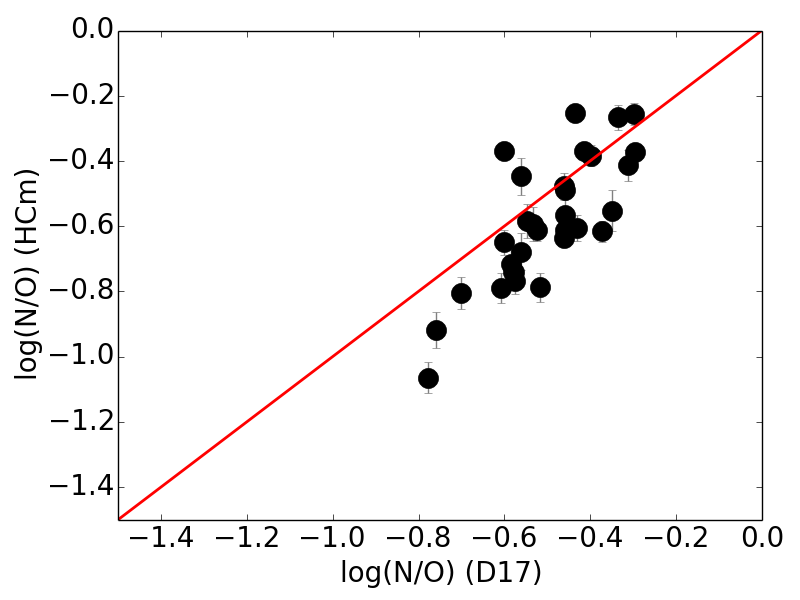}
\includegraphics[width=8cm,clip=]{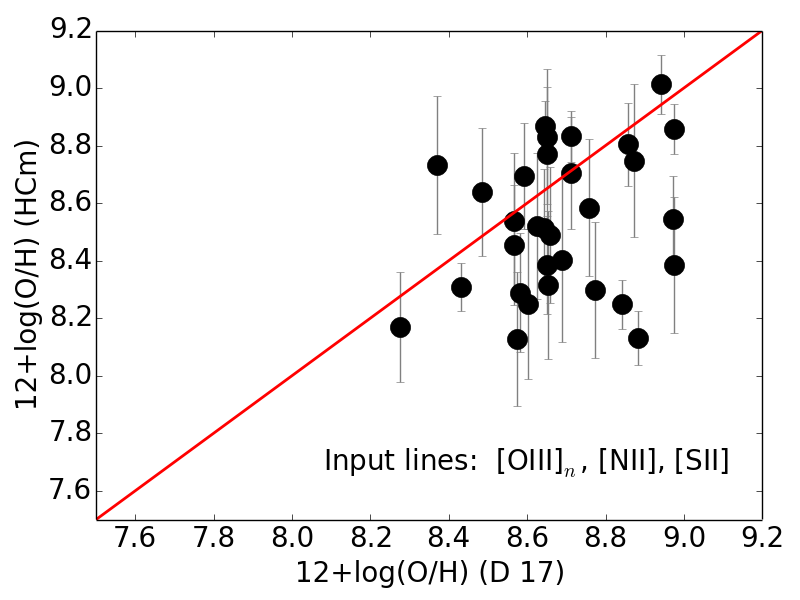}
\includegraphics[width=8cm,clip=]{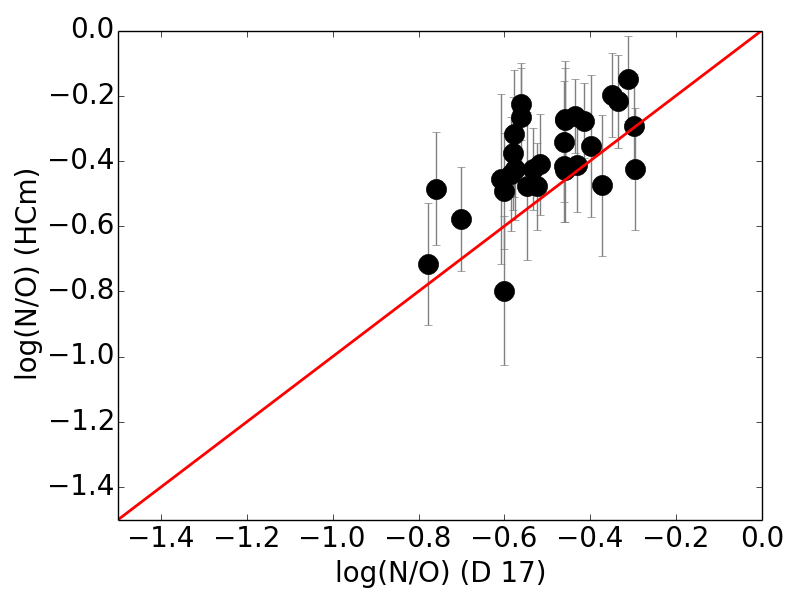}
\includegraphics[width=8cm,clip=]{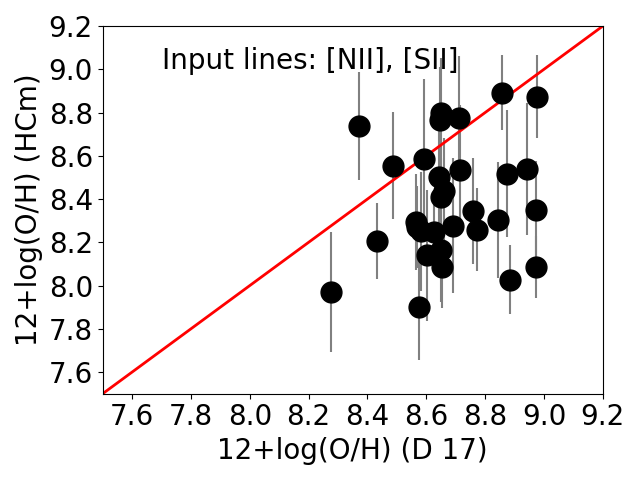}
\includegraphics[width=8cm,clip=]{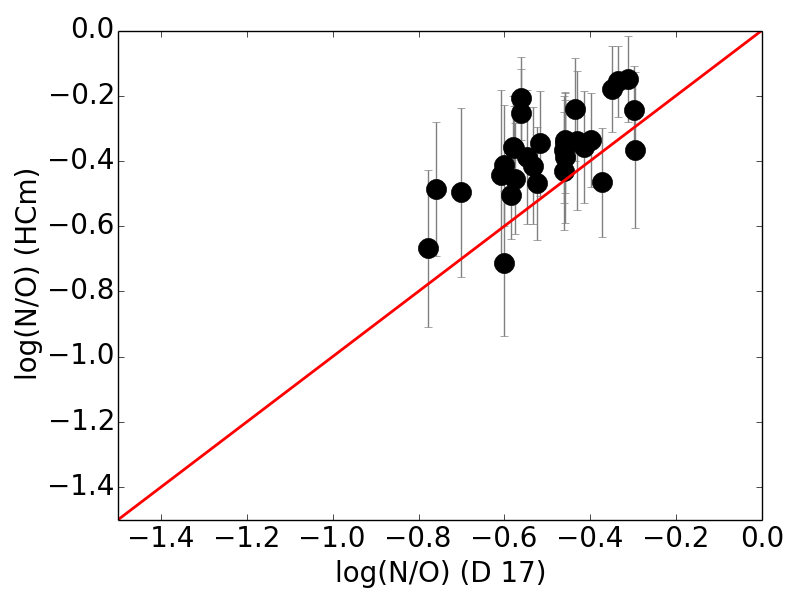}

\caption{Comparison between the chemical abundances derived from our method when only certain sets of lines are
considered and the abundances derived in Dors et al (2017) for the control sample. In top row, in absence of [\oii] $\lambda$3727 \AA; in middle row when [\oii]
and [\oiii] $\lambda$4363 \AA\ are not included and, in bottom row, when only [\nii] 6573 \AA\ and [\sii] $\lambda$$\lambda$6717,31 \AA\AA\ are included.
At left comparison of derived 12+log(O/H) and at right for log(N/O).
The red solid line represents the 1:1 relation.}

\label{comp_s4363}
\end{figure*}

When the [\oiii] $\lambda$5007/4363 \AA\ temperature-sensitive emission-line ratio cannot be measured in star-forming regions (or any other of the auroral-to-nebular  line-ratios
corresponding to other ionic species can be measured instead) implies that the so-called $T_e$ method (i.e. the calculation of all
chemical ionic abundances from a measured estimation of the electron temperature) cannot be used and other methods based only on strong emission-lines
can be adopted instead. In the case of the {\sc HCm} code
for star-forming regions, as described in \cite{hcm14}, this implies the adoption of an empirical relation between metallicity and
ionization parameter.
In the case of AGNs, though this dependence between $Z$ and $U$ is also found,, in absence of the [\oiii] $\lambda$4363 \AA\ lines there is no need to assume any extra relation to get accurate abundances
with the rest of strong lines, as there are not degeneracies between metallicity and the used optical 
emission-line ratios in the remaining grid of models for log $U$ $>$ -2.5.

In Figure \ref{comp_s4363} we show both O/H and N/O derived by our code for
the sample of analyzed galaxies when we just get a limited number 
of optical emission lines as compared with the abundances derived by \cite{dors17}. Using as input not all required lines  can be caused by a limited sensitivity or spectral coverage of our detector.
The mean offsets and standard deviation of the residuals are listed in Table \ref{dispersion}.

\begin{figure*}
\centering

\includegraphics[width=8cm,clip=]{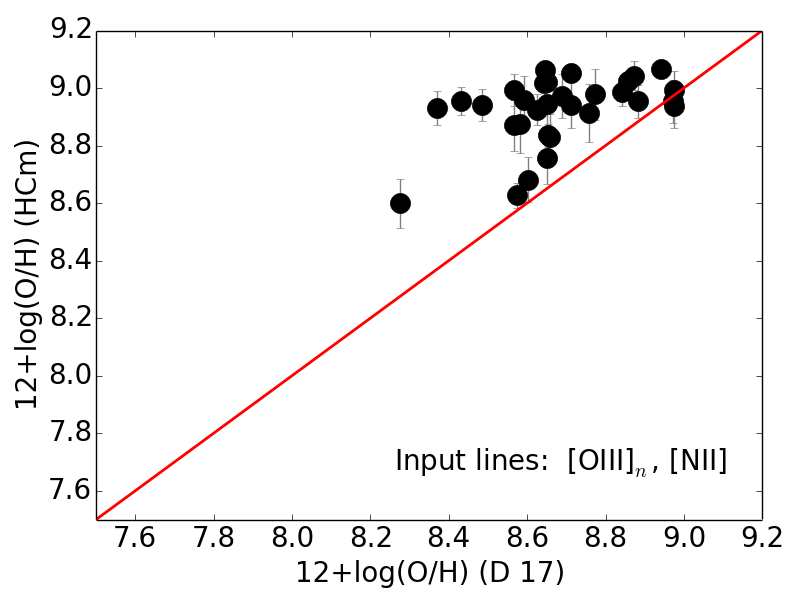}
\includegraphics[width=8cm,clip=]{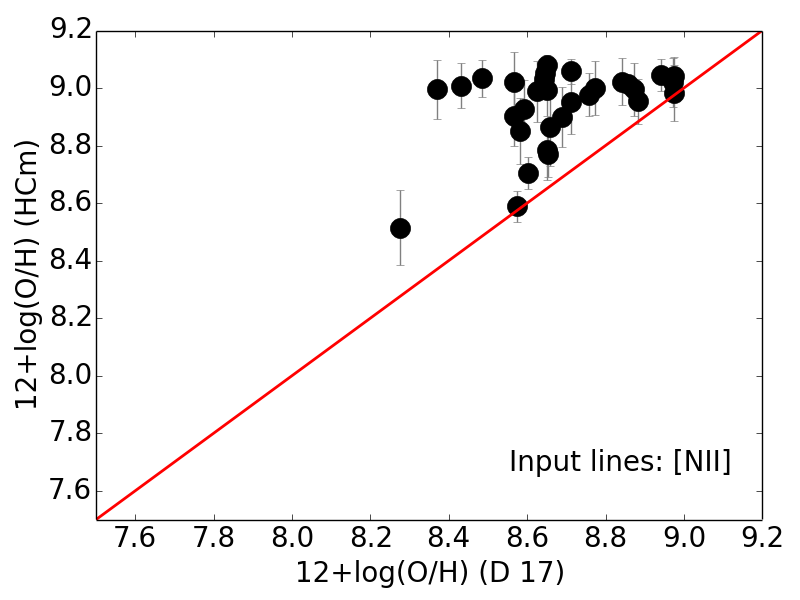}
\includegraphics[width=8cm,clip=]{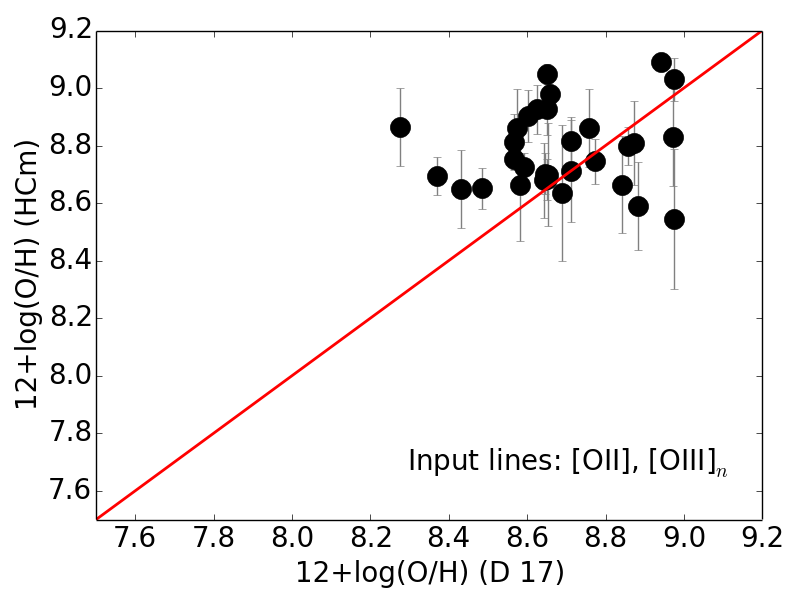}
\includegraphics[width=8cm,clip=]{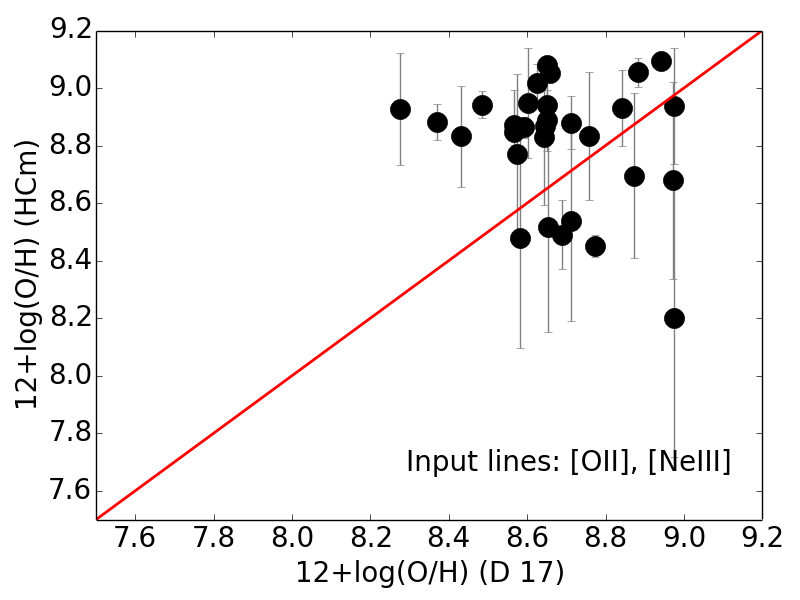}

\caption{Comparison between the oxygen abundances derived from our method when only certain lines are considered as input 
and a previous derivation of N/O is not carried out but an empirical O/H-N/O relation is adopted with respect to the values presented by Dors et al (2017) for the control sample. 
In the upper left panel using O3N2, in the right upper panel, using N2, 
in the lower left, using only [\oii] $\lambda$3727 \AA\ and [\oiii] $\lambda$5007 \AA, and, at lower right,
using [\oii] and [\neiii] $\lambda$3868 \AA.
The red solid line represents the 1:1 relation.}

\label{comp_r23}
\end{figure*}

In the case that no [\nii] $\lambda$6583 \AA\ emission line is detected in combination with another low-excitation line emitted
by an $\alpha$ element, such as [\oii] or [\sii], the code cannot calculate N/O.
In this case the code assumes the relation between O/H and N/O to follow the expected
relation derived from most star-forming regions and which can be seen in the Figure \ref{oh-no}. For lower values of metallicity N/O remains
constant and low due to that most of N production has a primary origin. As O/H enhances, at a value 12+log(O/H) $\sim$ 8.0, N/O
begins to
grow with metallicity as N has mostly a secondary origin. We assume for our models this restriction in
case N/O cannot be estimated, but this can produce deviations in those objects where this behavior
is not followed due to strong interactions with the intergalactic medium (e.g. \citealt{kh05}). In all case, this assumption is behind most
of the usual empirical and model-based calibrations based on [\nii] lines even if this relation has not been very well studied in the case of AGNs that can deviate
from the behavior observed in star-forming regions, so it is strongly preferable to have a good previous
estimation of N/O.

In Figure \ref{comp_r23} we show the comparison between the O/H values
derived by our code and those calculated by \cite{dors17} when no previous N/O is derived and a law between O/H and N/O is assumed instead,
for the case that we use as input lines O3N2 and N2. The mean offsets and the standard deviations of the
residuals are shown in Table \ref{dispersion}.

In case that no N/O previous calculation can be carried out, the code can also provides us with
a solution for O/H and $U$ using only [\oii] and [\oiii] 
 lines, or alternatively,
in case the observed wavelength range is limited and only very blue lines are observed, 
using only [\oii] and [\neiii] as input of the code, as discussed in hte previous sections
The comparisons between the total oxygen abundances obtained for these cases and those
from the control sample are shown in the lower panels of Figure \ref{comp_r23}. The corresponding means and deviations are also listed in Table \ref{dispersion}.

\subsection{Changing the resolution and range of the input grid}

\begin{figure*}
\centering

\includegraphics[width=8cm,clip=]{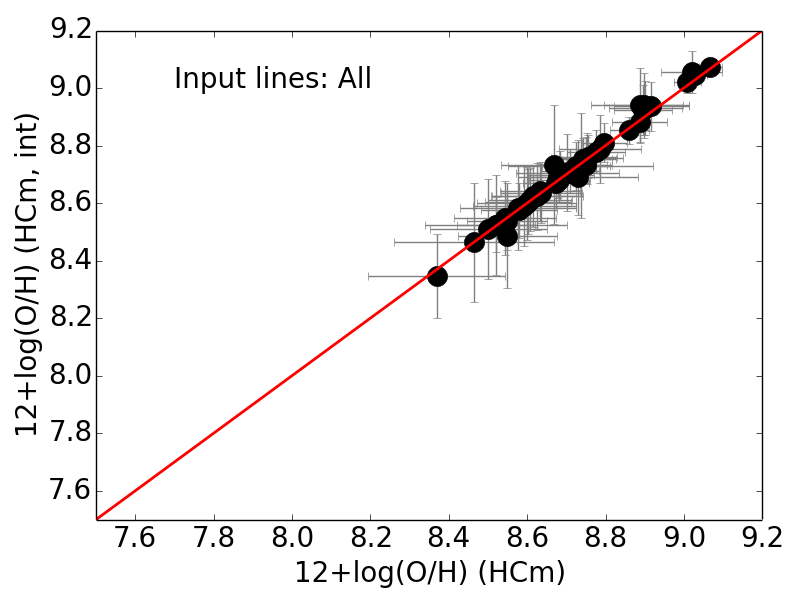}
\includegraphics[width=8cm,clip=]{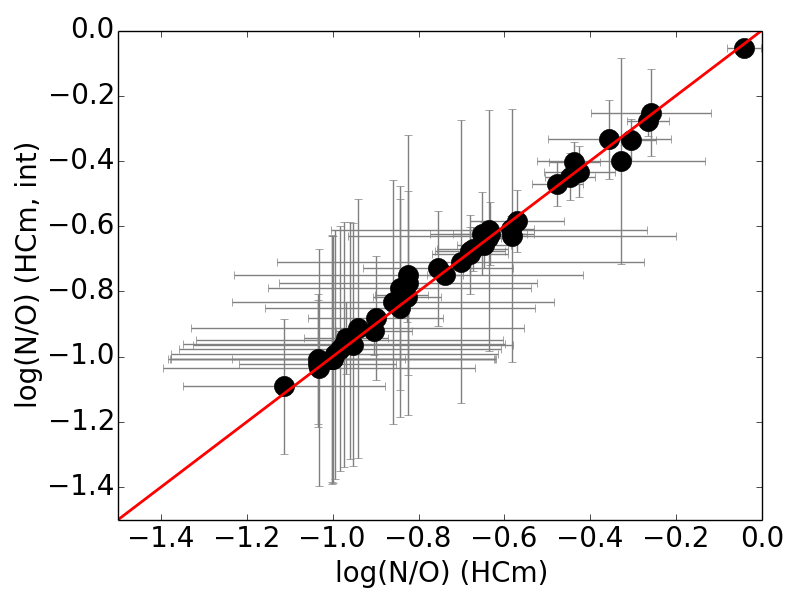}

\caption{Comparison between the resulting O/H (left panel) and N/O (right panel) derived by {\sc HCm}
using a non-interpolated and an interpolated grid of models. In both panels the red solid line represents the 1:1 relation.} 

\label{comp_int}
\end{figure*}

According to \cite{pm16} the use of a discrete grid of input values in {\sc HCm} can lead to a
somehow discrete distribution of the results that is above all noticeable for large samples of objects.
For this reason in the case of star-forming objects,  {\sc HCm} also supplies interpolations in the grid that multiplies in a factor 5 the resolution in the three main
input drivers of the grid (i.e. O/H, N/O, and $U$). In this version for NLRs of AGNs, this is also possible.
In Figure \ref{comp_int} we show the comparison between the results for O/H and N/O for the control sample when using the code with a
interpolated and a non-interpolated grid of models. As can be seen in both cases the agreement is very good and no deviations beyond the
obtained errors are produced (i.e. the standard deviation of the residuals is in both cases lower than 0.02 dex).

\begin{figure*}
\centering

\includegraphics[width=8cm,clip=]{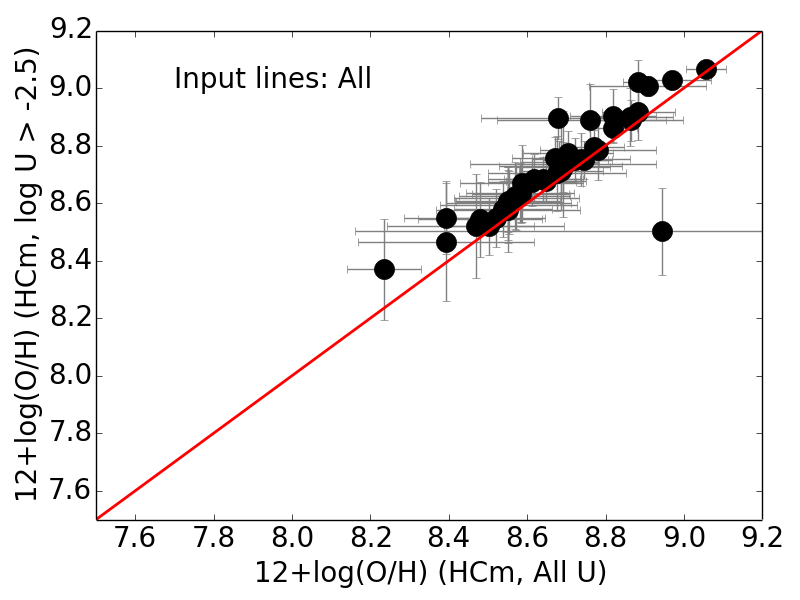}
\includegraphics[width=8cm,clip=]{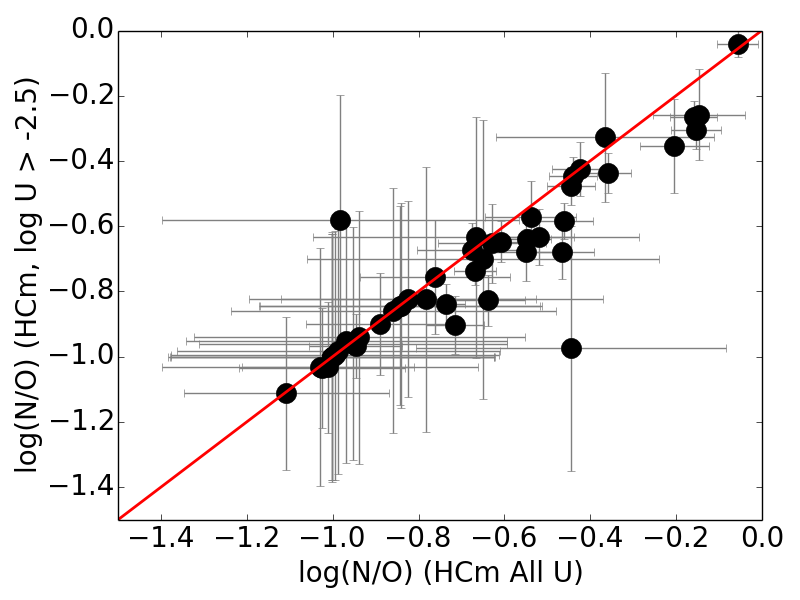}
\includegraphics[width=8cm,clip=]{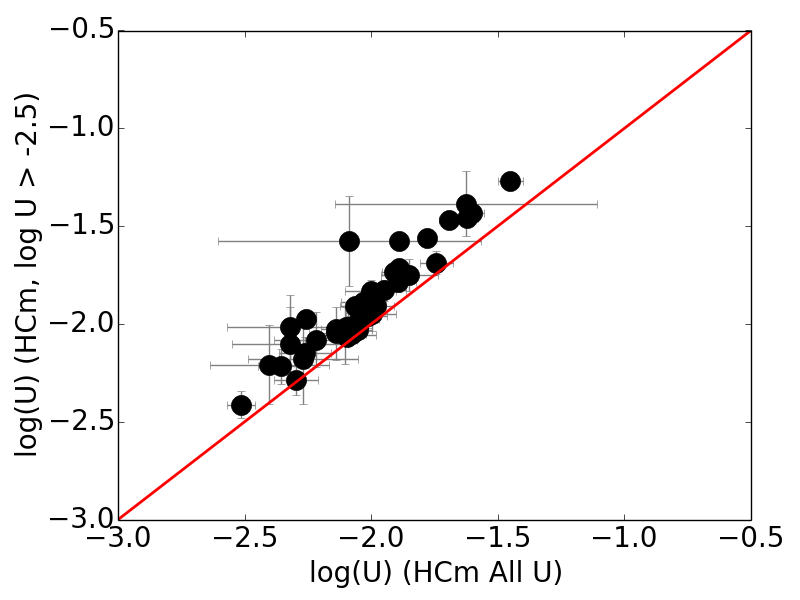}
\includegraphics[width=8cm,clip=]{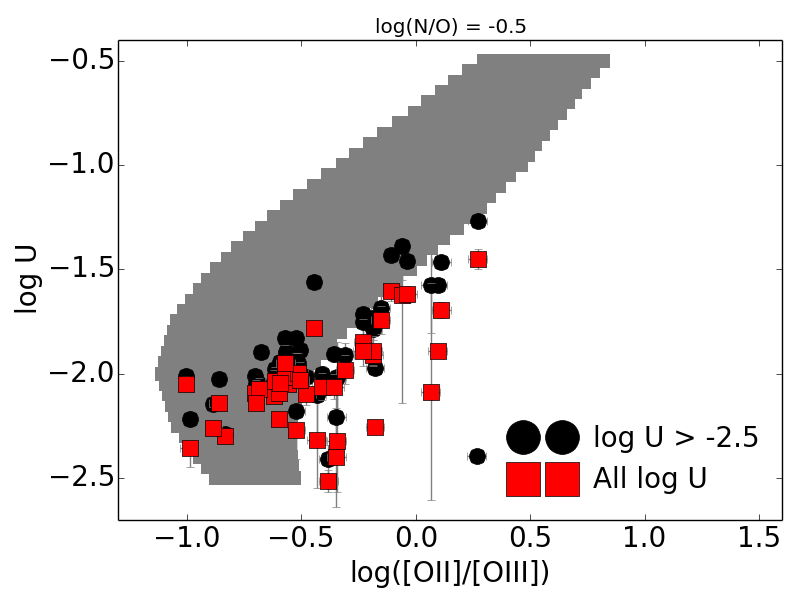}

\caption{Comparison between the resulting O/H ( upper left panel), N/O (upper right panel),
and log $U$ (left lower panel)  derived by {\sc HCm}
using a non-restricted and a $U$-restricted grid of models. In all these panels the red solid line represents the 1:1 relation.
The right lower panel represents log $U$ as a function of the  [\oii]/[\oiii] emission-line ratio
with models with O/H $\geq$ 8.4 as grey squares, black circles as results for the control sample in a $U$-restricted grid, and red squares as results for a non-restricted grid. } 

\label{comp_U}
\end{figure*}

Regarding the range of chosen input parameters in the models, abundances were selected to cover the expected values observed
in many NLRs and the extrapolation to the central regions in spiral galaxies. The fact that the code recovers the values found in \cite{dors17},
who do not consider any abundance restriction on their detailed search, probes that at least in this sample this range is appropriate.
In the case of the chosen restriction for log $U$ (i.e. only values log $U$ $>$ -2.5)  we explore to what extent this can affect 
the final derived abundances. In Figure \ref{comp_U} we show the comparisons between the abundances and $U$ obtained by the code when all
values in the grid are taken and when only the restricted range of $U$ is assumed. As it can be seen in the case of O/H and N/O this restriction does not
imply any difference within the errors. On the other hand the average log $U$ is 0.18 dex larger in average when the low values of $U$ are not taken, as can be
 expected as in the weighted means no models with very low values of $U$ are considered. Nevertheless, as can be seen also in the lower right panel of the same Figure, a large fraction of the objects whose $U$ was 
calculated using the entire range of $U$ lie in a region without any model that corresponds to the turnover region of the relation between
[\oii]/[\oiii] versus $U$. Although the choice of the upper branch in this relation can look arbitrary this is
the range expected for most AGNs (e.g. \citealt{dopita14}).

\subsection{Consistency of models with the $T_e$ method}

\begin{figure*}
\centering

\includegraphics[width=8cm,clip=]{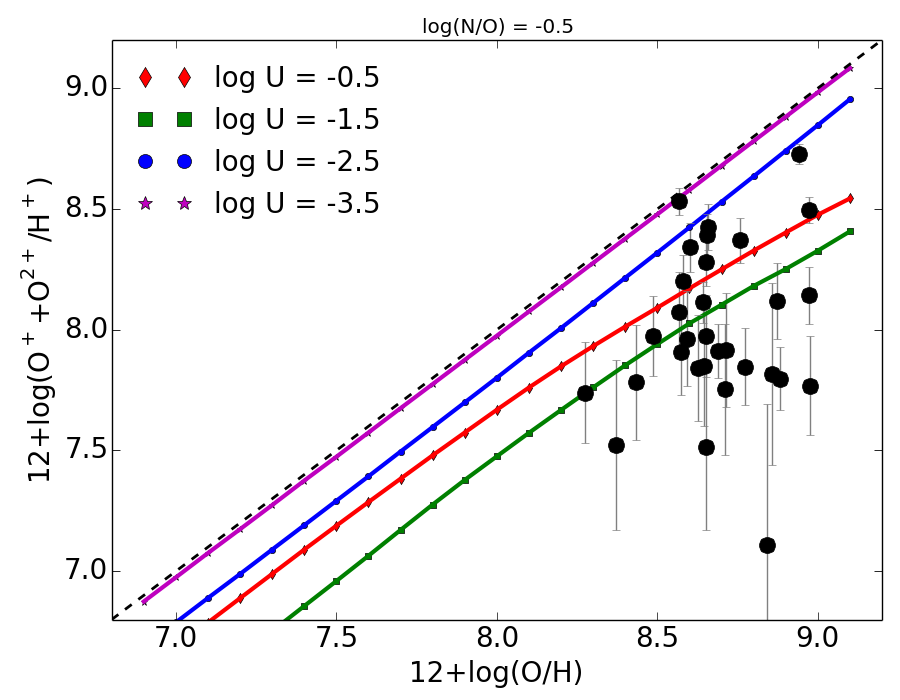}
\includegraphics[width=8cm,clip=]{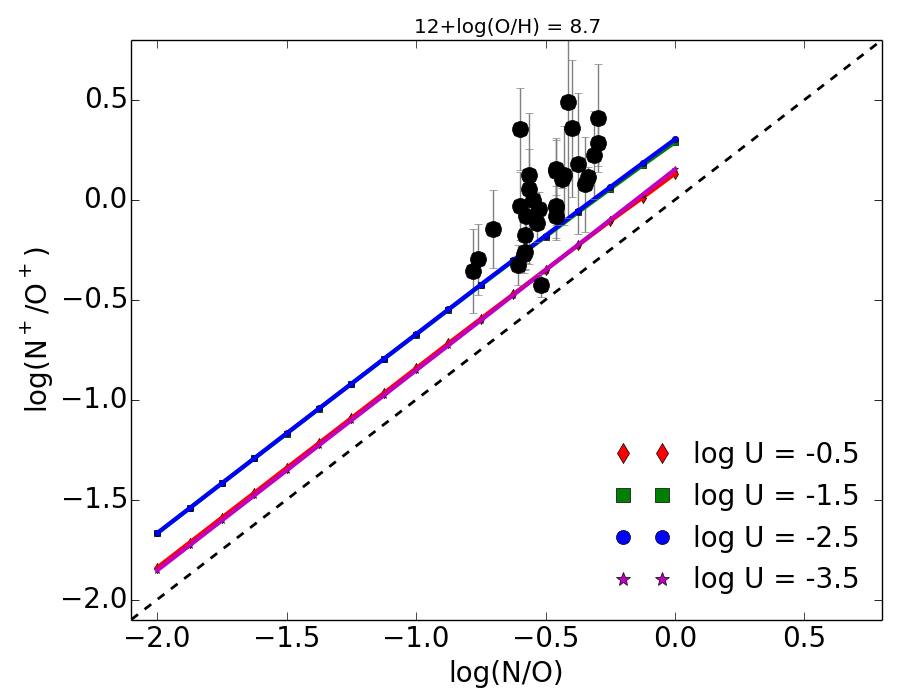}

\caption{Comparison between total oxygen abundance and the addition of the abundances of the two
main ions observed in the optical part of the spectrum (i.e. O$^+$ and O$^{2+}$), at left, and between N/O and the corresponding ion abundance ratio N$^+$/O$^+$, at right.
Models are represented using solid lines for different values of $U$.
In both panels the black circles represent the data from Dors et al (2017) whose total abundances where calculated
using tailored models, while their ionic abundances were calculated following the $T_e$ method.
The dashed black line represents the 1:1 relation.}

\label{icf}
\end{figure*}

It is known that the application of the $T_e$ method to derive total chemical abundances in the NLRs of type-2 AGNs can lead to
very low values as compared to those from photoionization models (e.g. \citealt{dors15}).
This discrepancy can be interpreted in terms of the usual offset that can be found between results from the $T_e$ method and models,
or to a different physics governing the ionization of the plasma. Since the code {\sc HCm} does not lead to 
any difference between its predictions and the application of the $T_e$ method in star-forming regions
 \citep{hcm14}, we can evaluate possible
explanations to this discrepancy using the version of the code for AGNs.

In Figure \ref{icf} we show the comparison between the total oxygen abundance derived by \cite{dors17}
for the compiled sample of Sy2 galaxies, compared to the O$^+$ + O$^{2+}$ relative to H$^+$ ionic abundances derived following
the $T_e$ method as described in \cite{tutorial}. This is based on  a previous determination of
the electron temperature from the emission-line ratio [\oiii] $\lambda$$\lambda$5007/4363 \AA\AA.
As can be seen, the objects lie in a range below the 1:1 relation as the addition of the relative abundance 
of O$^+$ and $O^{2+}$  is much lower than the total abundances derived
from models,  that is 0.7 dex in average.

We plot in the same figure the predictions made by the grid of models described in previous sections and we can see
that this difference is naturally explained by the models in terms of a large fraction of oxygen that it does not appear as the 
ions whose abundances can be derived using the optical emission-lines. According to models, the difference between the total oxygen abundance and the optical ionic
fractions depends strongly on total metallicity and ionization parameter and can reach up to 0.8 dex.

This difference underlines the importance of using models to derive the total abundance of oxygen in the NLRs of AGNs using optical lines as the
ionization correction factor (ICF) for O$^+$+O$^{2+}$, contrary to star-forming regions, is far to be negligible.

In the right panel of the same Figure we represent a similar relation comparing the total N/O ratio with the corresponding ionic
fraction N$^+$/O$^+$.
In this case this ionic fraction cannot be neither used as a proxy of N/O, as the empirical ionic ratio is in average
0.5 dex larger than the total elemental ratio. This offset is due to that the ionization structure of oxygen and nitrogen present
more differences than in the case of nebulae ionized by massive stars, and
N$^+$ represents a much larger fraction of N than O$^+$ of oxygen. In consequence, as in the previous case,
models are necessary to provide accurate ICFs to derive total chemical abundances from the observed ionic fractions using optical emission lines.

\subsection{Dependence on other input conditions}

\begin{figure*}
\centering

\includegraphics[width=8cm,clip=]{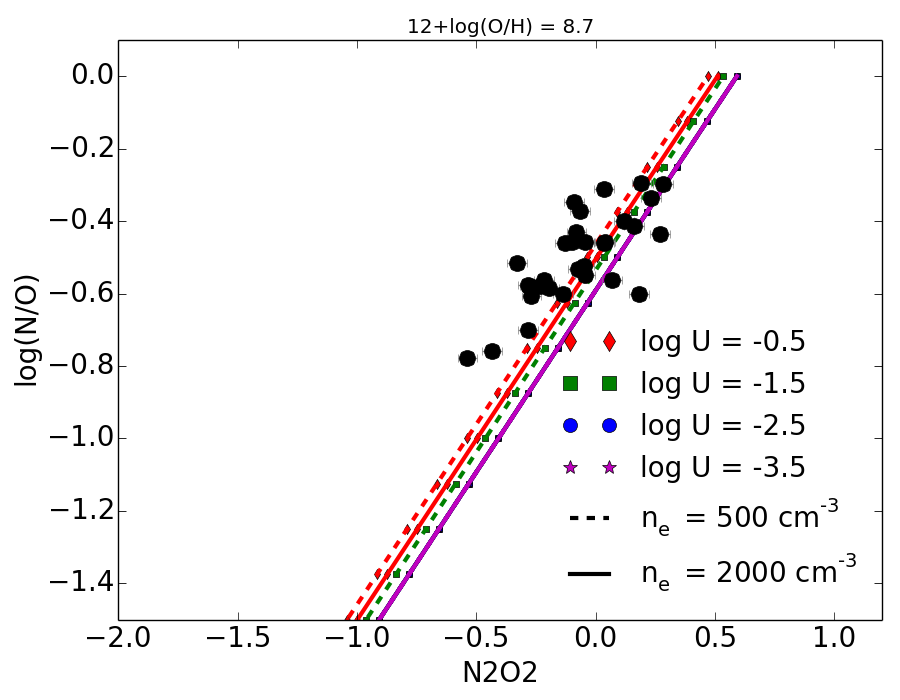}
\includegraphics[width=8cm,clip=]{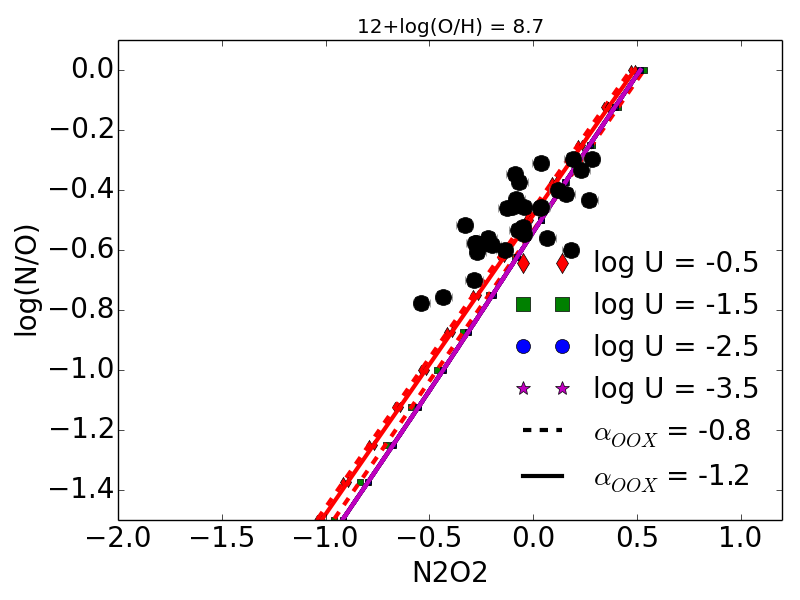}
\includegraphics[width=8cm,clip=]{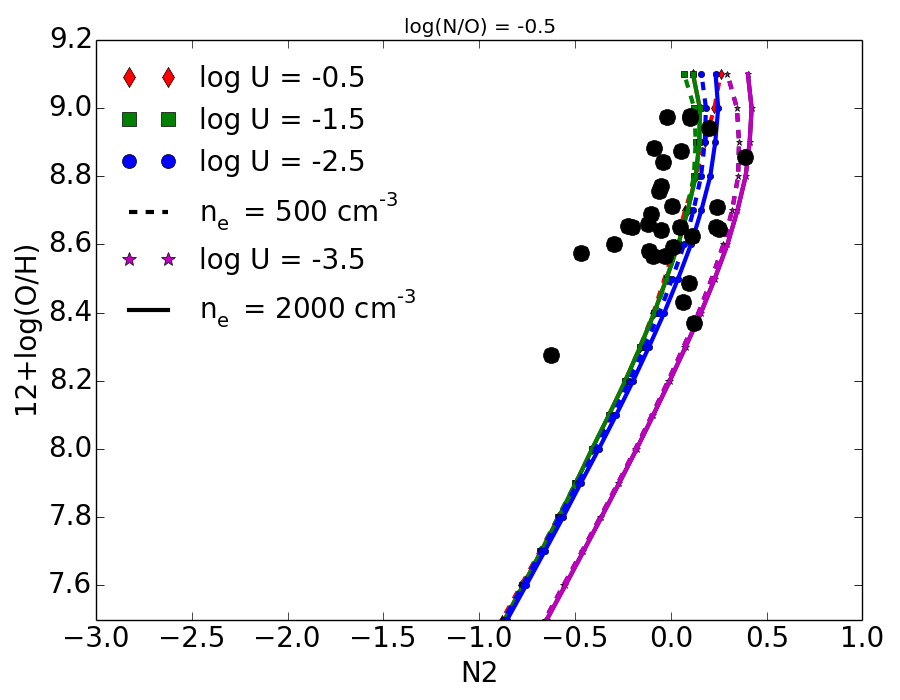}
\includegraphics[width=8cm,clip=]{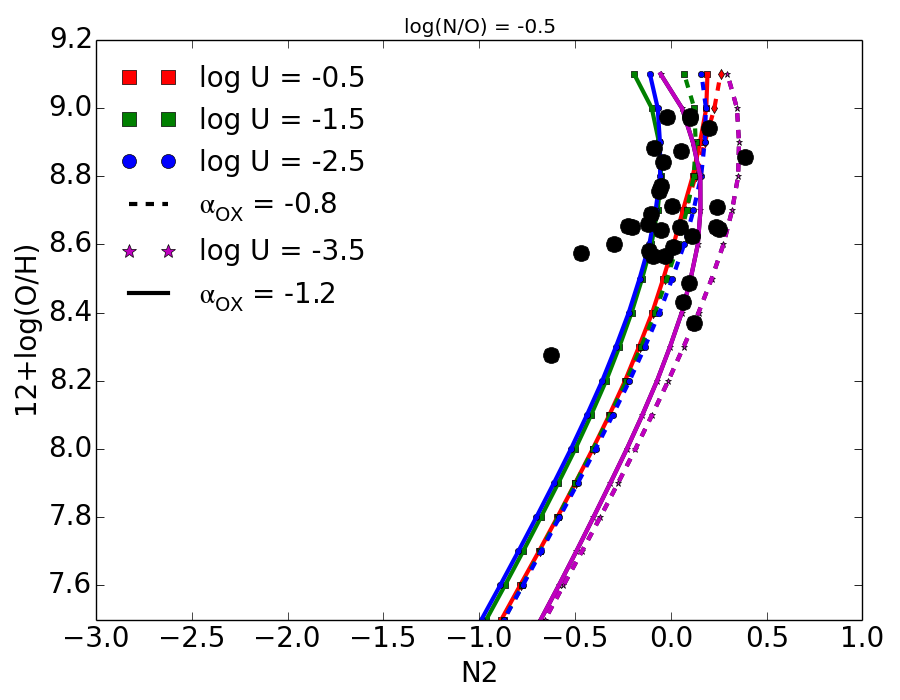}
\includegraphics[width=8cm,clip=]{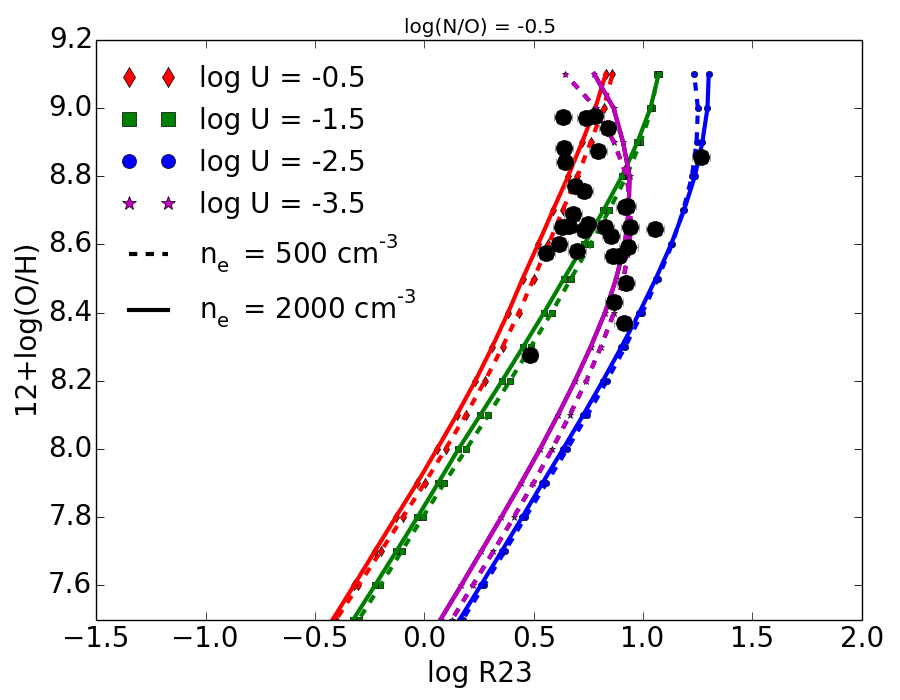}
\includegraphics[width=8cm,clip=]{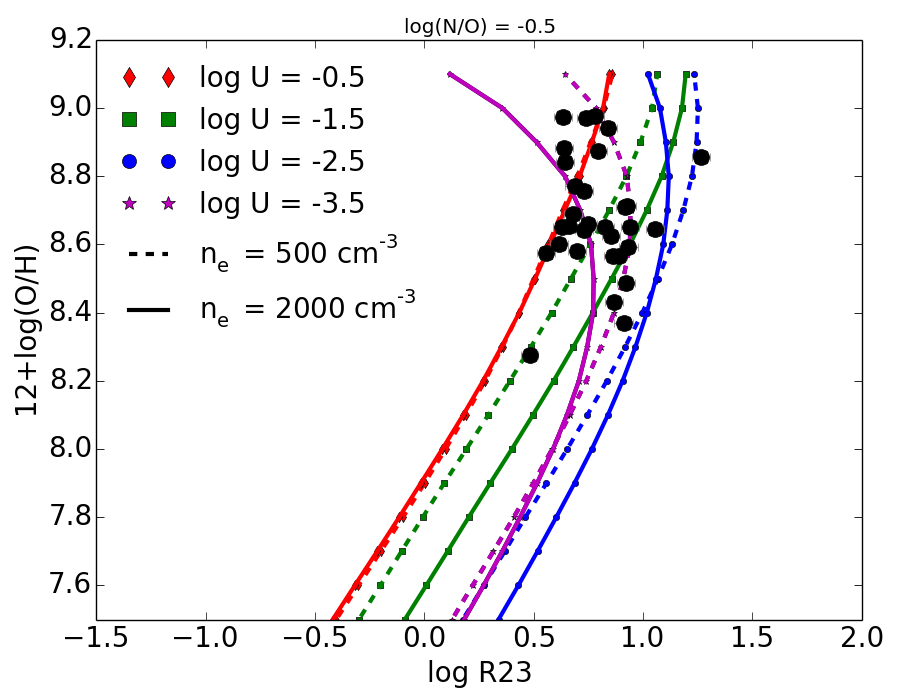}

\caption{Relations between different emission-line observables and abundances ratios used in the models as a function of different 
input parameters. Panels in the left column show the difference between models using an electron density of
2000 cm$^{-3}$ (solid line) and 500 cm$^{-3}$ (dashed line). The panels in the right column shows the difference between and $\alpha_{OX}$ for the input
ionizing SED of -1.2 (solid line) and -0.8 (dashed line). In the upper panels we show the relation between N2O2 and N/O for some models at a fixed 12+log(O/H) = 8.7, in the middle panels
the relation between N2 and O/H for some models at a fixed log(N/O) = -0.5 and in the bottom panels between R23 and O/H 
for some models at a fixed log(N/O) = -0.5.}

\label{obs_other}
\end{figure*}

\begin{figure*}
\centering

\includegraphics[width=8cm,clip=]{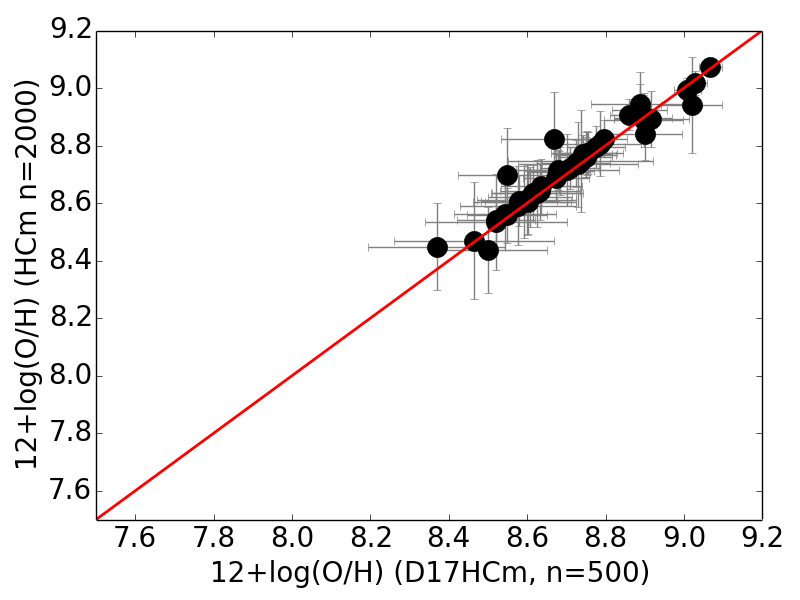}
\includegraphics[width=8cm,clip=]{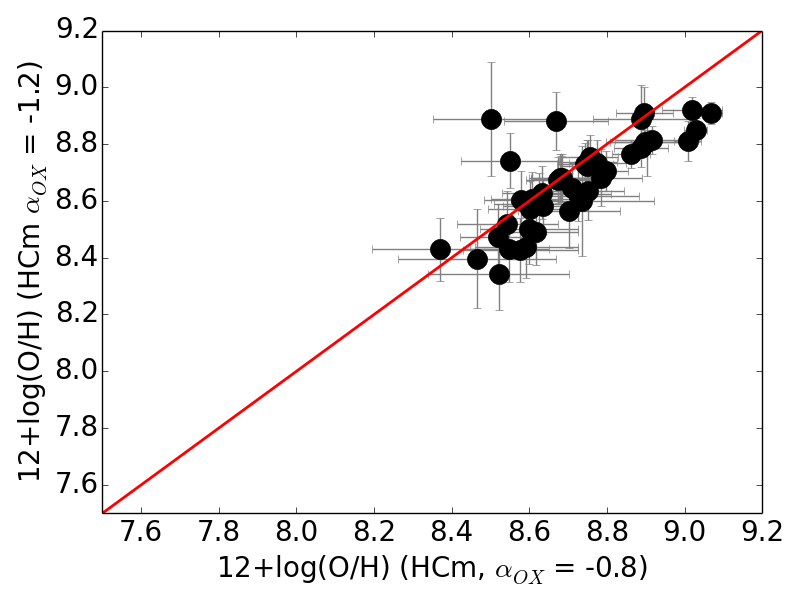}
\includegraphics[width=8cm,clip=]{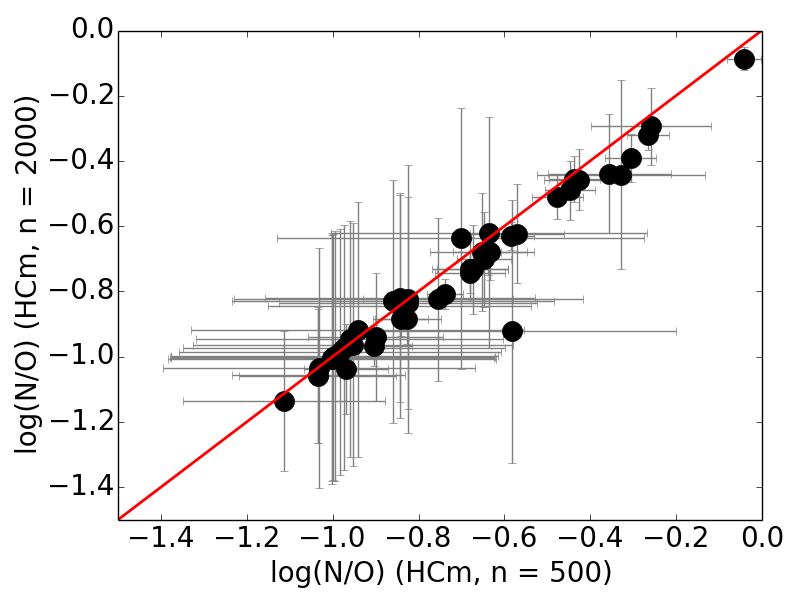}
\includegraphics[width=8cm,clip=]{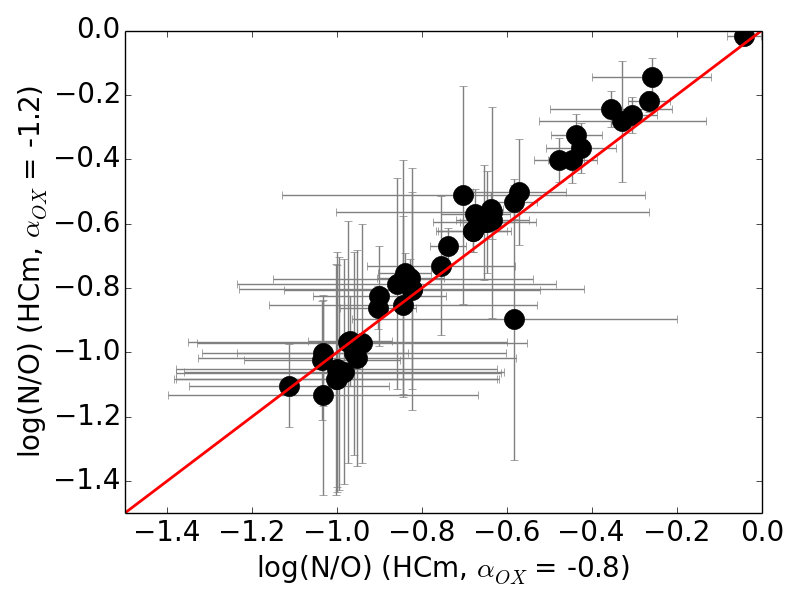}
\includegraphics[width=8cm,clip=]{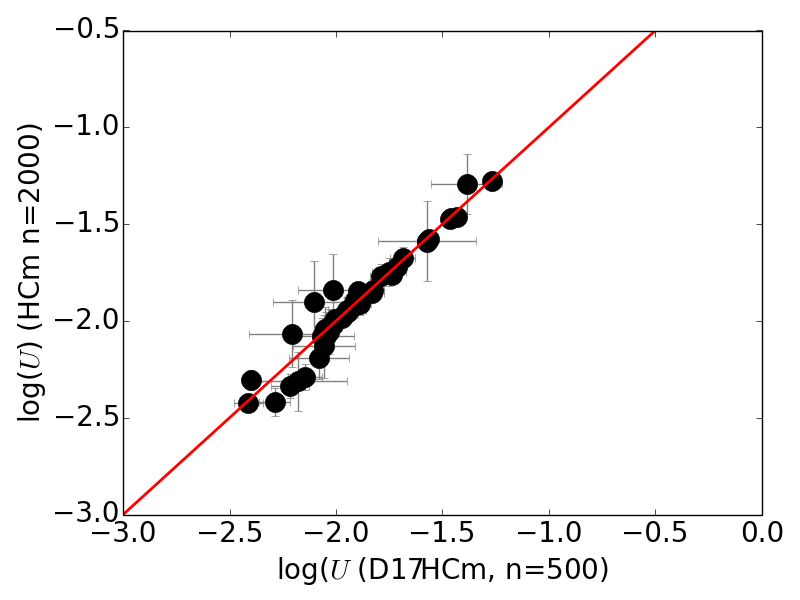}
\includegraphics[width=8cm,clip=]{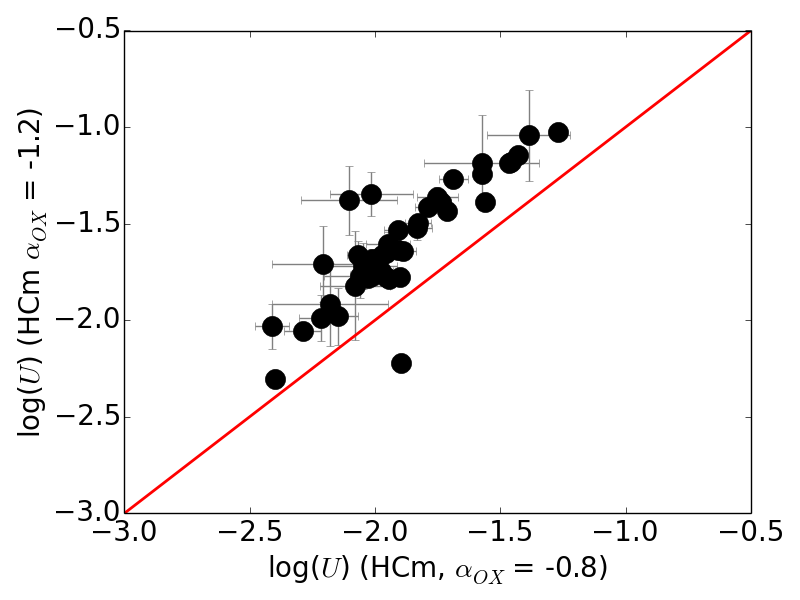}

\caption{Comparison between O/H (upper row), N/O (middle row) and log $U$ )lower row) 
from our method and all input emission lines using the grid of models with an electron density of 
500 cm$^{-3}$ and a $\alpha_{OX}$ = -0.8
and the results from the same code but changing two parameters in the input grid of models.
In left column changing density to n$_e$ = 2000 cm$^{-3}$ and in
the right column changing $\alpha_{OX}$ to -1.2.
The red solid line represents the 1:1 relation.}

\label{comp_other}
\end{figure*}

In this subsection we discuss how varying other input factors in the models can impact the
absolute values and uncertainties of the final derived abundances in our method.
Although there are multiple possible sources of uncertainty in the adopted input values in the models that can
affect our results in this paper we focus on electron density and the $\alpha_{ox}$ parameter as these are among 
those more difficult to be accurately estimated in NLRs of AGNs (e.g. \citealt{dorsuv}).

Electron density is one of the physical properties of the gas that can affect the emissivity of the observed emission lines.
Although it is known that below the critical density the collisional de-excitation is negligible 
in this case, it is important to quantify to what extent this can affect the final derived abundances.
In Figure \ref{comp_other} we show the comparison between the obtained O/H,  N/O, and log $U$ values derived by {\sc HCm} 
. When we changed the input electron density from 500 to 2\,000 particles per cm$^{-3}$  in the used grid of models.
As we can see no noticeable difference can be found in relation to the abundances or in $U$  derived using a lower density so the collisional de-excitation
does not imply a large difference.
Only in the case of N/O the change in the input density of the models imply values 0.04 dex lower for larger values of the
electron density.
Therefore in those cases where the electron density is lower than the critical density for the involved
emission-lines this is not going to be a key factor in the derived abundances.
However it is important to keep in mind that most of the times the
main source of information on the electron density is [\sii], a low-excitation ion, and this implies that if
an inner density structure exists in the NLR larger densities cannot be ruled out for high-excitation ions, so
a density diagnostics of this region is advisable.

We also show in the right panels of Figure \ref{comp_other} the differences obtained when wechange
the  $\alpha_{OX}$ parameter from -0.8 to -1.2.
In this case the change in the input conditions apparently does not affect much to the
derived O/H and N/O values..
Anyway the mean offsets are in all cases  lower than the typical obtained errors. The mean O/H abundances are in average 0.05 dex lower, while for
N/O we obtain values 0.03 dex larger.
On the other hand the change in she shape of the SED implies larger values of log $U$, which are 0.3 dex in average.
This shows that a consistent comparison of $U$ for objects with different radiation sources is much more difficult than in the case of chemical abundances.

\section{Summary and conclusions}

In the present work we have described a new methodology based on a bayesian-like comparison between the predictions from
a large grid of photoionization models and certain optical emission-line ratios to
provide estimations of the total oxygen abundances, the nitrogen-to-oxygen ratios, and
the ionization parameters in the NLRs of AGNs, along with uncertainties of these derived quantities depending on the
resulting $\chi^2$ distributions and the errors associated to observations.
This method has been included in the public version of the project {\sc Hii-Chi-mistry}
so it can be used consistently for large samples of objects.

The code firstly constrains the N/O ratio from the observational input taking advantage from the fact that this can
be derived using available ratios based only on low-excitation lines, such the N2O2 or the N2S2 parameter. The relations between
these two emission-line ratios and N/O appear to be different than that observed in star-forming objects owing to that a different ionization structure appears when the
ISM receives the radiation field from the central engine.

Once N/O is constrained, the code performs a second iteration through the space of models to derive O/H and $U$ using 
emission-line ratios similar to those used for star-forming objects, including RO3, N2, O3N2, R23 or O2Ne3. Again the behavior of these
result on very different relations with O/H or $U$, as an enhancement of the excitation and
a decrease of metallicity imply a lower fraction of O$^{2+}$ in the gas as much of oxygen appears as species more ionized.
This implies, for instance, that in this case R23 has a turnover in its  relation with O/H at a much higher values than for star-forming regions, while the relation between
the [\oii]/[\oiii] emission-line ratio with $U$ is double-valued. For this reason the code discards all those models with
log $U$ $<$ -2.5, although this restriction does not imply any significative change in the derivation of the resulting chemical abundances.

We compared the results of our code with the abundances derived using detailed photoionization models by \cite{dors17},
as no empirical derivation of chemical abundances using optical emission lines are available.
The analysis yields both O/H and N/O values very similar
to those obtained using tailor-made models. All galaxies belonging to the control sample lie in the high-metallicity regime in a range
8.37 $<$ 12+log(O/H) $<$ 9.07, and with a large production of
secondary nitrogen, as they lie in the range  -1.11 $<$ log(N/O) $<$ -0.04. However the objects present a large
dispersion in the O/H-N/O diagram that justifies a separate treatment of these two ratios in order to correctly use [\nii] emission lines to characterize the metal content of any object.

The mean ionization parameter for the whole sample is much larger (i.e. 1.5 dex)  than for a sample of pure star-forming
objects in the same metallicity regime with a very uncertain relation between $U$ and Z that should be studied in more detail.
However, contrary to star-forming regions, the code does not require 
any additional assumption on the O/H -$U$ relation in absence of certain lines to arrive to an acceptable solution in the calculation
of the chemical abundances.

We have checked if our method is consistent with the results obtained from the $T_e$ method in the derivation of
chemical abundances. It is known  that there is a large difference
between the total oxygen abundance derived from models and the results from the direct method in
NLRs of AGNs (e.g. \citealt{dors15}). Our results confirm that this difference is mainly due to the large fraction of oxygen that cannot be quantified by means
of the ionic fractions measured using optical emission-lines but a considerable amount of oxygen appears under the form of more ionized species, so
the use of models to derive a precise ionization correction factor is mandatory when other
wavelength regimes are not available.

Finally, we have also evaluated to what extent variations in some input conditions of the models imply deviations on
the obtained values by the code. In this way we have checked that assuming electron densities of the gas from 500 to 2 \,000 cm$^{-3}$ does not lead to
significant variations beyond the obtained errors. The same can be said when we consider an input SED with a parameter
$\alpha_{OX}$ going down from -0.8 to -1.2 but, in this last case, it cannot be ruled
out that this implies a larger variation in the obtained ionization parameter.

\section*{Acknowledgements}
This work has been partly funded by the Spanish MINECO
project  Estallidos 6 AYA2016-79724-C4.
and the Junta de Andaluc\'\i a for grant EXC/2011 FQM-7058. 
This work has been also supported by the
Spanish Science Ministry "Centro de Excelencia Severo Ochoa Program
under grant SEV-2017-0709.

EPM also acknowledges support from the CSIC intramural grant 20165010-12
and the assistance from his guide dog Rocko without whose daily help this work would have been much more difficult.
RGB acknowledges support from the Spanish Ministerio de Econom\'ia y Competitividad, through 
projects AYA2016-77846-P and AYA2014- 57490-P. RGB acknowledges support from the Spanish Ministerio 
de Econom\'ia y Competitividad, through project AYA2016-77846-P.




\bibliographystyle{mnras}

\bibliography{HCm-AGN} 








\bsp	
\label{lastpage}
\end{document}